\documentclass{aa}  
\newcommand{\degree}{$^{\circ}$\,}
\newcommand{\mstar}{$M_{\star}$}
\newcommand{\lrad}{$L_{150\,\rm MHz}$\,}
\usepackage{graphicx}
\usepackage{tabularx}
\usepackage{txfonts}
\begin{document}

   \title{Radio AGN jet alignment}

   \subtitle{Radio and Optical Position Angle of Radio Galaxies}

   \author{
          Xuechen Zheng\inst{1,2}
          \and
          Yuze Zhang\inst{2}
          \and
          Huub R\"ottgering\inst{2}
          }

   \institute{Key Laboratory for Research in Galaxies and Cosmology, Shanghai Astronomical Observatory, Chinese Academy of Sciences, 80 Nandan Road, Shanghai 200030, China\\
   \email{zhengxc@shao.ac.cn}\\
   \and 
   Leiden Observatory, Leiden University,
              PO Box 9513, 2300 Leiden, RA, The Netherlands\\
             }

   \date{Received ; accepted }

 
  \abstract
{ It is well established that active galactic nuclei (AGNs) play an important role in the evolution of galaxies.  
These AGNs can be linked to the accretion processes onto massive black holes and past merger events in their host galaxies, which may lead to different alignments of the jets with respect to the host galaxies. 
   This paper presents a study of the position angle (PA) differences between radio and optical images of radio AGNs based on the second data release (DR2) of the Low Frequency Array (LOFAR) Two-Meter Sky Survey (LoTSS), the {\it Karl G. Jansky} Very Large Array Faint Images of the Radio Sky at Twenty-Centimeters Survey (FIRST), the Dark Energy Spectroscopic Instrument (DESI) Legacy Imaging Surveys and the Sloan Digital Sky Survey (SDSS). 
   We assessed PA measurement biases in the data and classified the radio AGNs based on the radio luminosity and infrared colour from the Wide-field Infrared Survey Explorer (WISE). 
   This resulted in the largest yet published sample of 3682 radio AGNs with reliable radio and optical PA measurements.
   The PA difference (dPA) distributions for the radio AGN sample show a prominent minor-axis alignment tendency. 
   Based on some simple assumptions, we simulated the projection effect to estimate the intrinsic jet-galaxy alignment.
   The observed dPA distribution can be well described by a two-component jet-alignment model in which one component is more aligned with the minor axis of the host galaxy than the other.
   The fitting results indicate that the jet alignment is dependent on radio luminosity and the shape of the host galaxies, with the jets being more likely to be aligned with the minor axis of the galaxy for lower radio luminosity and for optically more elongated radio AGNs.
   The minor-axis alignment of the entire sample may suggest a coherent accretion model present in most AGN host galaxies, while a considerable number of luminous radio-AGN with massive host galaxies might have undergone an accretion according to the chaotic model or past merger events.
    }


   \maketitle
%

\section{Introduction}

    It has been known for a long time that active galactic nuclei (AGNs) play an important role in the evolution of their host galaxies \citep{2009ApJ...690...20S}. These AGNs are the result of the accretion process of the super-massive black holes (SMBHs) at the centre of galaxies. Multiple evidence has shown that an increase in inflow leads to the prevalence of AGN activity \citep[e.g.][]{Best05a,Dunn06,Best07}. AGNs release considerable energy into the surrounding galactic medium and the inter-galactic medium \citep[IGM; ][]{1993MNRAS.264L..25B}. The energy released by AGNs takes two different forms: radiation and mechanical energy \citep{Best12}. The energy output has been related to the classification of AGNs, by comparing it with the corresponding accretion rate at the Eddington limit \citep[Eddington rate; ][]{Best12}. Radiation is dominant for high excitation AGNs, which are a type of radiative mode AGNs with strong high-ionization narrow lines \citep{Best12}. These sources, along with the rest of the radiative mode AGN population, are products of efficient accretion at a rate above 1 per cent of the corresponding Eddington rate. The mechanical energy of ejected particles is dominant for low-excitation AGNs with weak narrow low-excitation emission lines, which results from a lower accretion rate \citep{1979MNRAS.188..111H,1994ASPC...54..201L}. Regardless of the form, the energy outflow could potentially disrupt the inflow of gas and cease the AGN activity. 

    AGN jets from radio observations provide a probe of the accretion of SMBHs. There are currently thought to be two types of inflow in the central region of host galaxies: the coherent model and the chaotic model \citep{2011MNRAS.414.2148L}. The coherent model is based on angular momentum preservation of the inflow, while in the chaotic model, there are shifts in angular momentum throughout the inflow process \citep{2011MNRAS.414.2148L}. Revealed by previous simulation results, different inflow mechanisms result in different SMBH spin alignments and, therefore, radio AGN jet directions \citep{2012MNRAS.425.1121H,2011MNRAS.414.2148L}. Under the conservation of angular momentum,  jets from AGNs described by a coherent accretion model are mostly perpendicular to their host galaxies' galactic plane. By contrast, jets from AGNs following a chaotic accretion model are randomly orientated with respect to their host galaxies. On top of these accretion models, an AGN jet might indicate possible merger events. During a merger event between two galaxies the spin of SMBHs can change significantly \citep{2008ApJ...684..829E,2011ApJ...741L..33B}. Also, the resulting AGN jet orientation is expected to be different after the merger.
    
    Despite the dichotomy of these simulation results, the two models, as well as possible merger events, suggest a link between AGN jet orientations and the accretion process. This relationship can be studied by comparing the radio and optical counterparts of the same source.
    The orientation of radio jets with respect to the host galaxies has been studied since 1970s.
    The earliest investigation by \citet{Mackay71} suggested that the major axes of radio sources were aligned with the major axes of their optical counterparts based on 18 sources.
    However, later studies using larger sample indicated the radio major axes were either aligned with optical minor axes \citep{Palimaka79,Guthrie79,Guthrie80,Birkinshaw85,Sadler89,Condon91,Andernach95} or not correlated with the optical major axis \citep{Gibson75,Sullivan75,Valtonen83,Sansom87}.  
    In the recent observation-based study  conducted by \citet{2009MNRAS.399.1888B} based on over 14000 galaxies from the {\it Karl G. Jansky} Very Large Array (VLA) Faint Images of the Radio Sky at Twenty-Centimeters Survey \citep[FIRST;][]{Becker95} and the Sloan Digital Sky Survey \citep[SDSS; ][]{York00}, the radio-loud sub-sample showed random orientations with respect to their hosts.
    They also suggested a possible bimodality in the radio-optical misalignment angle distribution of AGNs.
    However, this sample is contaminated by star-forming (SF) galaxies with the bulk of radio emissions emerging from star formation inside the galactic plane. 
    The most recent investigation on the jet orientation was done by \citet{VazquezNajar19} based on over 2000 extended radio galaxies, which indicated an optical minor axis alignment of radio jets and no evidence for a bimodality.
    The contradictory results in these studies could be due to the limited sample sizes or possible bias in the sample selections.
    This bias could be related to the dependence of jet orientation on sizes, galaxy shapes or types of jets \citep{Andreasyan84,Andreasyan99,Andernach09}.
    Moreover, the projection effect, the discrepancy between intrinsic alignment and apparent alignment, might also hide the intrinsic alignment information.
    
    In this paper, we present an analysis of the misalignment angle between optical and radio images of radio AGNs based on the largest sample used so far with reliable position angle (PA) measurement, along with simulations which demonstrate the link between the observed and intrinsic radio jet alignment with respect to their host galaxies. The primary goal of this study is to provide a further understanding of the accretion process for radio-loud AGN and their triggering/fueling mechanisms.
    
    In this work, we mainly use the data from the Low Frequecy Array (LOFAR) Two-metre Sky Survey \citep[LoTSS;][]{Shimwell22}. LOFAR reaches more than an order of magnitude deeper than the FIRST survey for sources with a typical spectral index \citep{vHaarlem13}. The survey also has higher sensitivity to extended radio structures \citep{Sabater19}. Hence, it is ideal for studying radio jet morphology in this research. The remaining sections of this paper are as follows. We first describe the data and data reduction process in Sect. \ref{data}. In Sect. \ref{roalign}, the result of the radio-optical misalignment angle distribution of the overall sample and sub-samples based on different divisions is presented. Next, in Sect. \ref{atoi}, we showcase the simulation results which link the apparent alignment result from observation to the intrinsic alignment of AGN jets. Finally, in Sect. \ref{dis}, we discuss various physical implications of our observation and simulation results. 
    Throughout this paper, we adopt a cosmology with the following relevant parameters: $\rm{H_0 = 70 kms^{-1} Mpc^{-1}}$, $\rm{\Omega_{m} = 0.3}$, $\rm{\Omega_{\Lambda} = 0.7}$.

\section{Data and sample}
\label{data}
In this work, we combine the radio data from the LoTSS DR2 \citep{Shimwell22} and the FIRST survey \citep{Becker95} to select radio AGNs with reliable PA measurements.
In the following section, we introduce the bias in the dataset and the selection process to construct a reliable sample.
\subsection{LoTSS DR2}
Our radio sample is first selected from the LoTSS DR2 \citep{Shimwell22}.
The LoTSS project aims to observe the entire northern hemisphere in the frequency band 120-168 MHz using LOFAR.
The LoTSS DR2 covers 27\% of the northern sky (5634 square degrees) with an rms noise level of $83\,\rm\mu Jy\,beam^{-1}$ and a resolution of 6$\arcsec$.
The area of coverage of LoTSS DR2 is the combination of two contiguous regions centred at RA $\sim13$h and RA $\sim1$h respectively.
A total of 4\,396\,228 radio sources were detected in the Stokes I maps.
The point-source completeness of this survey was about 90\% at a peak brightness of 0.8$\,\rm mJy\,beam^{-1}$ and the positional accuracy was about 0.2$\arcsec$.

In the LoTSS DR2, the source detection process was performed using the Python Blob Detector and Source Finder \citep[PyBDSF;][]{Mohan15} on the mosaic images.
PyBDSF uses wavelet decomposition to find high signal-to-noise (S/N) peaks in the images and fits them with one or more Gaussian components.
These Gaussian components were grouped automatically to form radio sources in the LoTSS DR2 source catalogue.
This process could detect and associate small sources ($\lesssim15\arcsec$) with simple structures in the LoTSS DR2.
The PyBDSF results provide the deconvolved size and position angle for the small and simple radio sources, contributing to the estimation of the jet alignment in the following analyses.
The association of larger sources was processed by the Radio Galaxy Zoo: LOFAR Zooniverse project ('RGZ(L)' hereafter)\footnote{\url{https://www.zooniverse.org/projects/chrismrp/radio-galaxy-zoo-lofar}}.

The optical data used in the cross-matching process was based on the Dark Energy Spectroscopic Instrument (DESI) Legacy Imaging Surveys (hereafter the Legacy Surveys).
The Legacy Surveys were a combination of three public projects, the Dark Energy Camera Legacy Survey, the Beijing-Arizona Sky Survey and the Mayall $z$-band Legacy Survey, using three telescopes: the Blanco telescope at the Cerro Tololo Inter-American Observatory; the Mayall Telescope at the Kitt Peak National Observatory; and the University of Arizona Steward Observatory 2.3 m (90 inches) Bart Bok Telescope at
Kitt Peak National Observatory.
A total of 14\,000 square degrees of extragalactic sky in the northern hemisphere was observed in the $grz$ bands with uniform depths ($\sim$23 AB magnitude at $r$ band).
The point-spread functions (PSFs) in the Legacy Survey have a median full width at half maximum (FWHM) of about 1$\arcsec$, comparable to that for the SDSS images \citep{Abazajian09}.
The software package {\it Tractor} \citep{Lang16} was used to extract sources from the stack images and construct the source catalogue with morphological information, including the shapes and profiles of the galaxies. 
Based on this catalogue of optically detected objects, the Legacy Surveys also provided forced-photometry mid-infrared magnitudes for each source in four bands (3.4, 4.6, 12 and 22 $\mu$m, also known as the W1, W2, W3 and W4 bands) using the stack images from the Wide-field Infrared Survey Explorer \citep[WISE][]{Wright10,Cutri13} and NEOWISE-Reactivation \citep[NEOWISE-R][]{Mainzer14}.

    \begin{figure*}
    \centering
    \resizebox{1\textwidth}{!}{\includegraphics{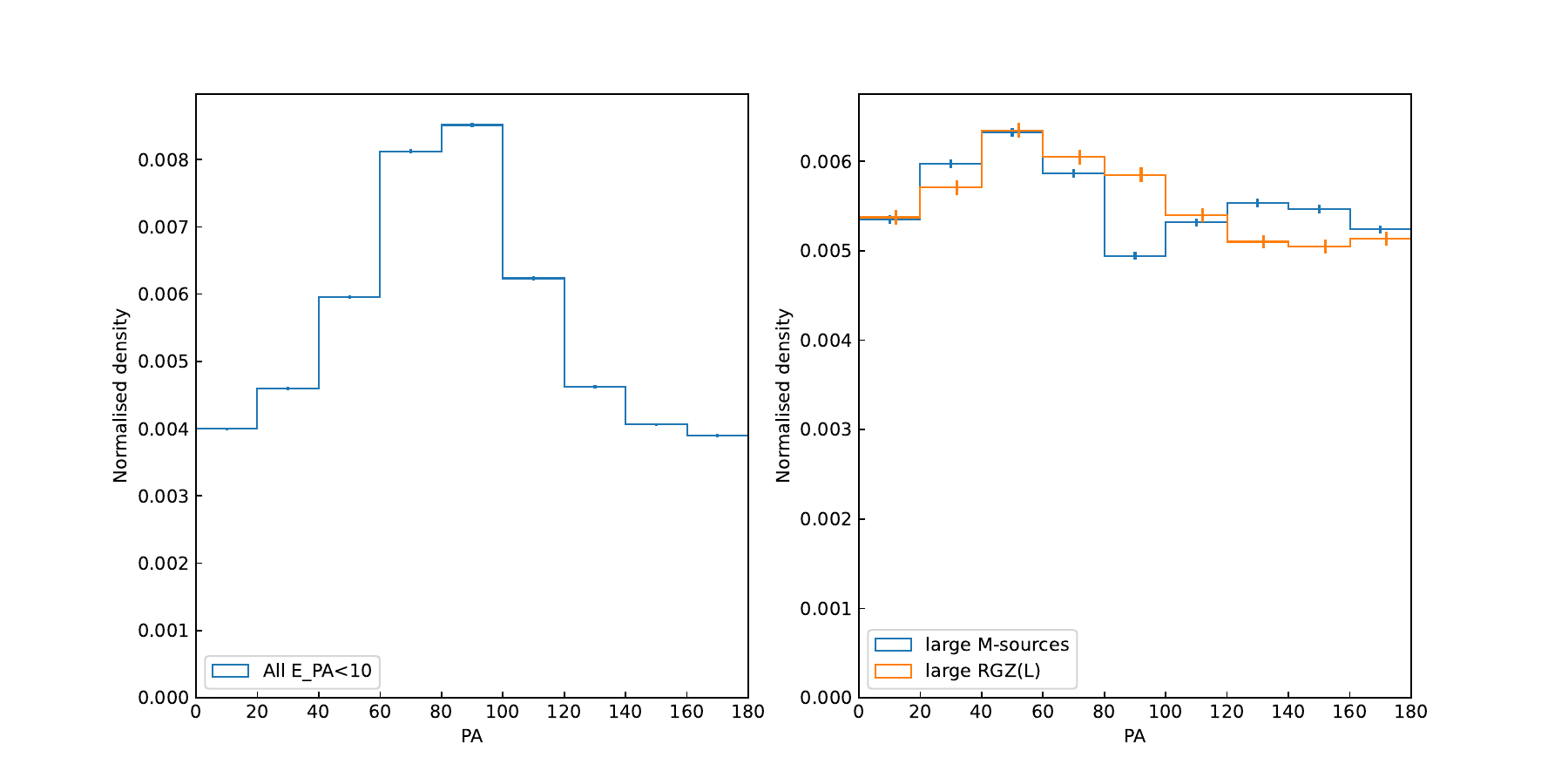}}
    \caption{Normalised distribution of RPAs in bins of 20\degree for the radio sources in the LoTSS DR2.
    The RPAs are based on the results from \citet{Hardcastle23}.
    The panel on the left shows the distribution of RPAs of radio sources in the LoTSS DR2 with PA errors smaller than 10 \degree. 
    The panel on the right shows the distribution of RPAs of the large multiple-Gaussian sources (noted as M-sources) and the large RGZ(L) sources. 
    The blue histogram shows the RPA distribution of the large multiple-Gaussian sources with a valid deconvolved RPA and the red histogram shows the distribution of the large RGZ(L) sources.
    The two histograms are normalised to make the area of integration equal to one.
    The errorbars are derived from Poissonian errors.}

    \label{fig:PA_dist}       
    \end{figure*}
The PyBDSF catalogue was refined and cross-matched with optical/infrared data following  procedures similar to those used by \citet{Williams19} to construct the value-added-catalogue (VAC) for the LoTSS DR2 \citep{Hardcastle23}.
The cross-matching procedures combined the likelihood ratio (LR) cross-matching method \citep[e.g.][]{Sutherland92} and visual inspections.
For each isolated and small radio source, the probability of a nearby optical object (candidate) being its counterpart was estimated based on the position, brightness and colours of the object.
This probability was then compared with the probability of the optical object being a random interloper, resulting in the LR of the optical source.
The optical counterpart for the radio source is the candidate with the largest LR unless the LR was below a defined threshold.
For the large radio sources, the Gaussian component associations and the optical counterpart identifications were done by citizen scientists and radio astronomy experts in the RGZ(L) \citep[see ][for more details]{Hardcastle23}.
Based on the cross-matching procedures, about 70\% of the radio sources detected in the LoTSS DR2 have an optical counterpart in the Legacy Surveys.
%
%
Of the sources in the LoTSS DR2 VAC, 57\% have a reliable spectroscopic or photometric redshift.
The spectroscopic redshifts are from the SDSS data release 16 \citep[SDSS DR16][]{Ahumada20}, early release of the DESI spectroscopic survey \citep{DESI23}, the HETDEX data release \citep{Mentuch23} and the spectroscopic redshifts from \citet{Gloudemans22}.
The photometric redshifts are based on the results from \citet{Duncan22}.
The sources in this work are all selected from these radio sources with an optical counterpart.

\subsection{Sample selection}
To study the jet alignment within the galaxies, it is important to have reliable position angle measurements in both optical and radio bands.
In this section, we describe the bias and uncertainties in the measured radio PA (RPA) and optical PA (OPA) in the data, and the criteria we use to select a sample with reliable PAs.
All the PAs in the following sections are defined in the range $[0,180)$ degree and measured from north through east.

\subsubsection{RPA in the LoTSS DR2}\label{sec:rpa}
The RPA in the LoTSS DR2 VAC was estimated in different ways depending on the source type \citep{Williams19}.
The RPA of radio sources with only one Gaussian component (`S\_Code'=`S' in the VAC) were defined by the major axis of the Gaussian component.
For radio sources fitted by multiple Gaussian components in PyBDSF and not processed by the RGZ(L), which have a `S\_Code'=`M', the RPA was determined by moment analysis.
For the radio sources associated with the RGZ(L), the RPA was the PA of the largest diameter vector of the source.
The three methods have consistent and reliable results for large, simple-structure sources.
However, the measured RPA appears to be significantly biased for small sources.
In Fig. \ref{fig:PA_dist}, we show the deconvolved RPA distribution of all the radio sources with a valid deconvolved RPA measurement\footnote{The size of the restored beam in LoTSS DR2 \citep{Shimwell22} is 6\arcsec$\times$6\arcsec. A source with the minor axis smaller than 6\arcsec\, would not have reliable deconvolved measurement including `DC\_PA' in the LoTSS DR2 VAC.} and a small RPA error (`E\_DC\_PA'$<10$\degree in the LoTSS DR2 VAC). 
While the RPA of sources in the sky is expected to be uniformly distributed within 0\degree to 180\degree, the RPA distribution of the radio sources in the LoTSS DR2 has a prominent peak around 90\degree.
We note that this bias is likely related to the actual beam shape in the LoTSS DR2.
However, given the complexity and uncertainty in the LOFAR radio image processing, we did not investigate further as it was not our focus here.
%
%
As this bias is only significant for small sources, we only included large sources that were not affected in this work. 

\begin{figure*}
    \centering
    \resizebox{1\textwidth}{!}{\includegraphics{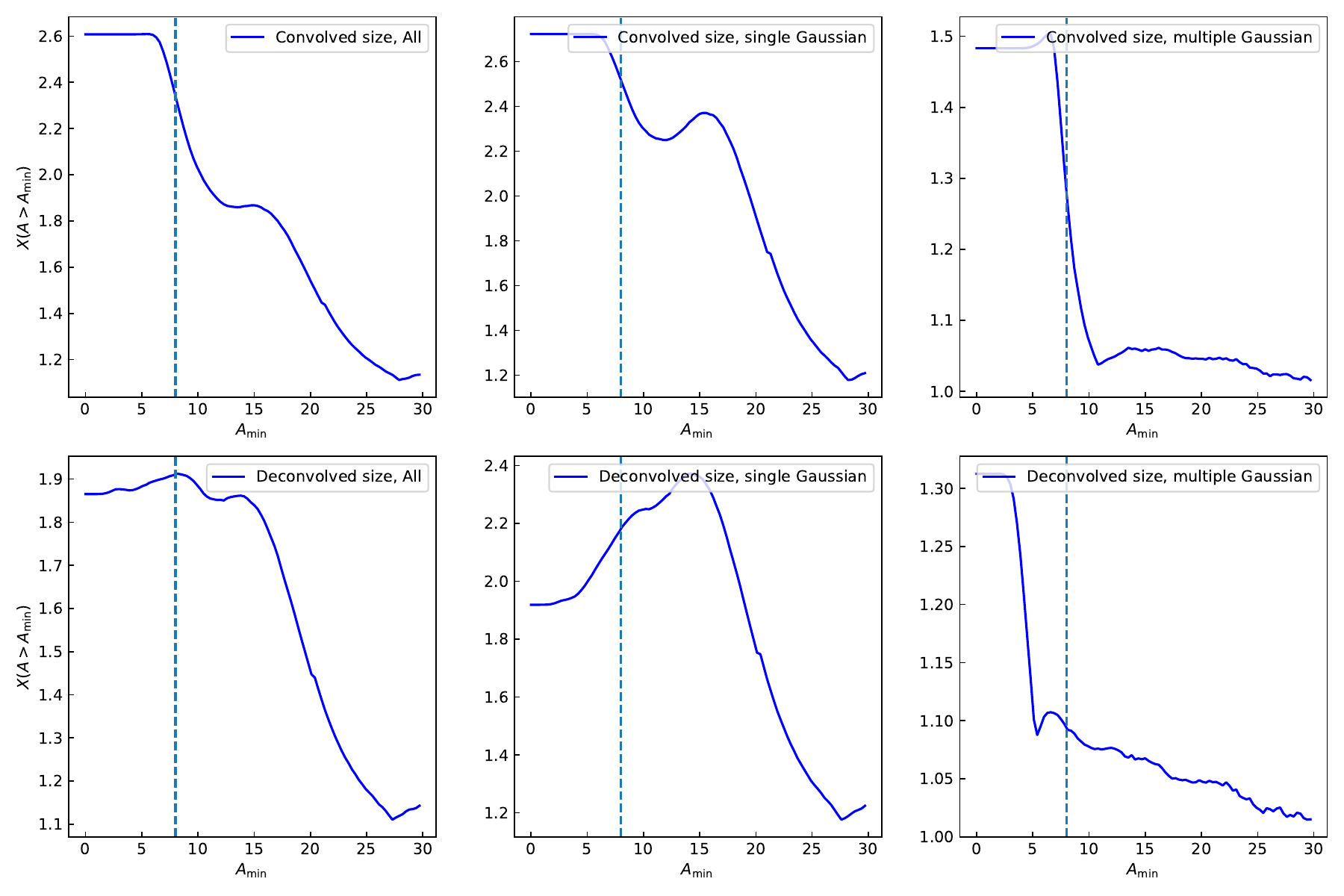}}
    \caption{The peak excess $X(A>A_{\rm min})$ of radio sources in the LoTSS DR2 VAC with different minimum source angular sizes $A_{\rm min}$(convolved or deconvolved major axis).
    Top panels: the peak excess as a function of the minimum convolved sizes.
    Bottom panels: the peak excess as a function of the minimum deconvolved sizes.
    The peak excess for all radio sources, single Gaussian radio sources and multiple Gaussian radio sources are shown from left to right respectively.
    The dashed lines in each panel denote a minimum size of 8\arcsec.}
    \label{fig:peak_excess}       
\end{figure*}

To investigate which sources are large enough to be free from the 90\degree bias, we calculated the number of sources with RPA$=60-120$\degree, $n_{\rm peak}(A>A_{\rm min})$, and outside 60--120\degree, $n_{\rm outside}(A>A_{\rm min})$, for radio sources with an angular size $A$ larger than $A_{\rm min}$.
We defined the `peak excess' as
\begin{equation}
   X(A>A_{\rm min})=2n_{\rm peak}(A>A_{\rm min})/n_{\rm outside}(A>A_{\rm min}) 
\end{equation}
and show the peak excess as a function of the minimum source size $A_{\rm min}$ (`Maj' or deconvolved size `DC\_Maj' in the LoTSS DR2 VAC) in Fig. \ref{fig:peak_excess}.
The peak excesses for all radio sources, single Gaussian radio sources and multiple Gaussian radio sources are listed from left to right respectively in Fig. \ref{fig:peak_excess}.

It is clear that the 90\degree\, bias is more significant for small and single Gaussian sources.
The source density at the peak RPA can be $\sim$1.5 to 2.6 times larger than the non-peak RPA regions for radio sources with an $A_{\rm min}$ of less than 8\arcsec.
For single Gaussian sources, the peak excess only drops to 1.2 for sources with an angular size larger than $\sim25\arcsec$.
The peak excess is less influential for multiple Gaussian sources as it drops to less than 1.1 at a minimum deconvolved angular size of $\sim5\arcsec$. 

Based on the peak excess results, we only select multiple component sources and  with a deconvolved angular size larger than 8\arcsec\, to have an unbiased RPA measurement.
The RPA distributions of these sources are shown in the right panel of Fig. \ref{fig:PA_dist}.
The 90\degree bias is not significant in these sources, but the distribution of RPA still differs from a uniform distribution.
This deviation from uniformity may be related to the residual effect of the beam shape bias or an alignment of jets of unknown origin \citep[e.g.][]{Contigiani17,2020A&A...642A..70O}.
We note that this deviation is less than 10\% different from the average value and will therefore likely not influence our analyses significantly.

In addition, we also require the radio sources to be a resolved source in the LoTSS DR2 VAC. 
In LoTSS DR2, a source is marked as `unresolved' when its measured major axis size is less than 15\arcsec\, and the ratio of integrated flux density ($S_{I}$) to the peak brightness ($S_{p}$) $S_{I}/S_{p}$ is within the 99.9\% percentile of the $S_{I}/S_{p}$ 
 distribution of the genuine point-like sources with similar S/N selected in \citet{Shimwell22}.
This means that a source is more likely to be a point source rather than a source with extended structures \citep[for more details, see][]{Shimwell22}.
In this case, the shape of the radio source would be close to the beam shape and the RPA measured by moment analysis would not indicate the RPA of the physical radio source.

We also only selected simple radio sources with well-defined RPAs.
In LoTSS DR2 VAC, the catalogue RPAs for LR-detected sources are estimated by moment analyses taking all the components into account.
For the RGZ(L) sources, the catalogue RPAs are the PAs of the largest diameter vector determined by the convex hull containing all the components of the sources \citep[for more details, see][]{Williams19,Hardcastle23}.
These estimates work well for simple sources with all components basically aligned in one direction, e.g. simple two-sided jets.
However, for sources with extended and complex structures, such as with X-shape or Z-shape or blurred features, the catalogue RPAs might not be a good indicator of jet direction because the measured RPAs might vary largely with different methods and definitions.
To select the radio sources with simple structure, we calculated the component-determined RPA (RPA$_{\rm comp}$) for each source.
If a radio source has a simple structure, the RPA$_{\rm comp}$ would be consistent with the catalogue RPA.
Based on the differences between RPA$_{\rm comp}$ and catalogue RPA, $\Delta$RPA, we only included sources with $\Delta$RPA<20\degree in the following analyses.

In summary, we adopted the following criteria to select radio sources with an unbiased and well-defined RPA:
\begin{itemize}
    \item[(1)] the LR-detected sources with the measuring uncertainty of deconvolved RPA `E\_DC\_PA'$<$10\degree, or the RGZ(L) sources that are not likely to be blended or problematic sources (`Blend\_prob' or `Other\_prob' $<0.2$ in the LoTSS DR2 VAC);\footnote{The RGZ(L) sources do not have an error estimation for RPA, but they are typically large and verified by human eyes. Therefore we assume they have negligible measurement errors. Instead, the possibility that they are blended sources or sources with other problems is used to evaluate the reliability of the measurement.}
    \item[(2)] the deconvolved angular size `DC\_Maj'$>8\arcsec$ or `Composite\_Size' (size determined by the RGZ(L) ) $>8\arcsec$;
    \item[(3)] multiple Gaussian components sources (`S\_Code'=`M') or RGZ(L) sources (`S\_Code'=`Z');
    \item[(4)] keyword `Resolved'=`True' in the LoTSS DR2 VAC;
    \item[(5)] $\Delta$RPA<20\degree.
\end{itemize}

\subsubsection{Optical positional angles}\label{sec:opa}
\begin{figure}
    \centering
    \includegraphics[width=\linewidth]{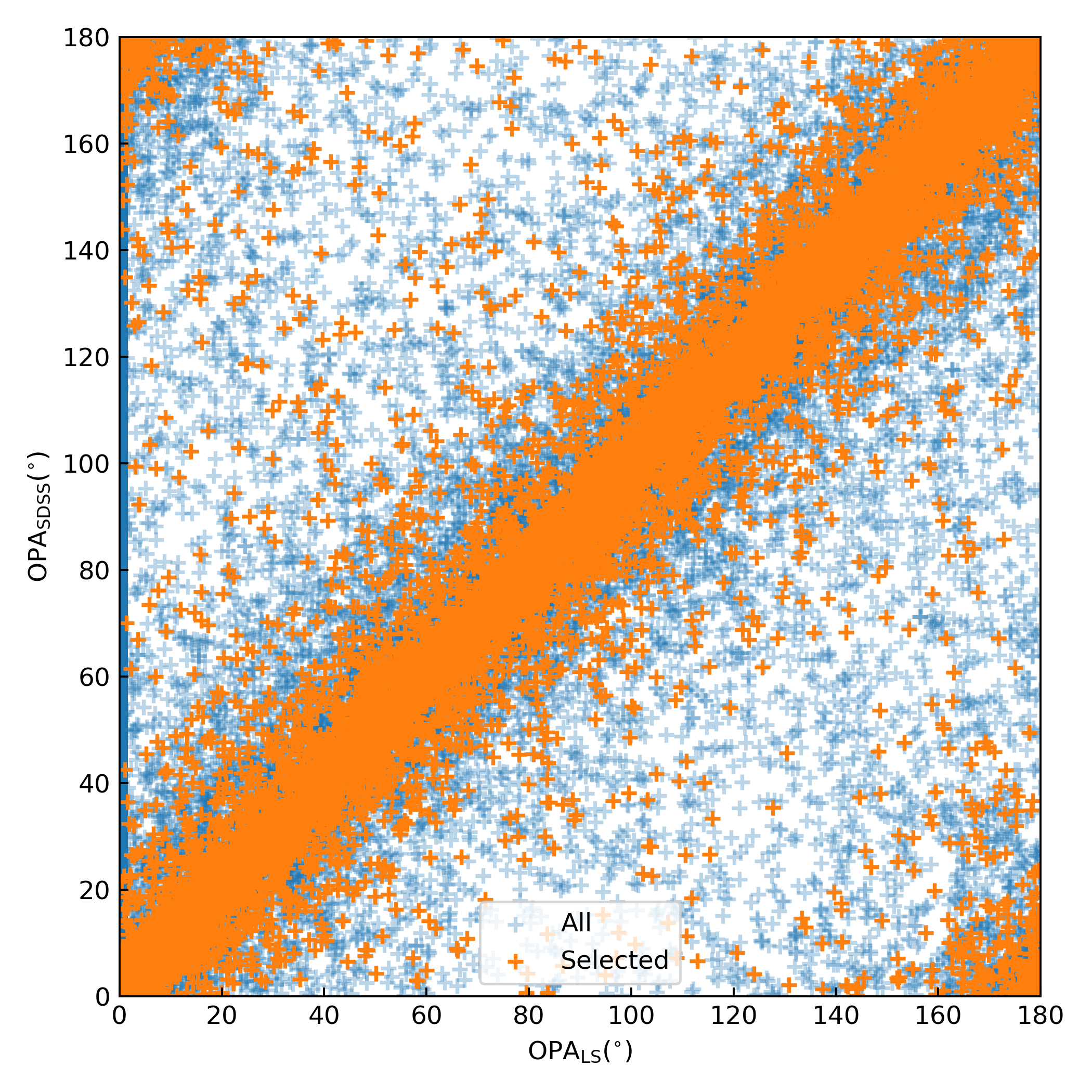}
    \caption{The OPAs of sources with detections in both the Legacy Surveys and the SDSS in this work.
    The x-coordinates denote the OPAs in the Legacy Surveys and the y-coordinates denote the OPAs in the SDSS.
    The blue symbols are all the LS-SDSS sources before adopting criteria (6)-(8).
    The orange symbols the LS-SDSS sources selected by criteria (6)-(8).
    The orange symbols are overplotted on top of the blue symbols.
    The two clusters in the upper left and lower right corners also correspond to the consistency of OPAs in the Legacy Surveys and the SDSS, because of the continuity of PA measures below 0\degree and above 180\degree .
    In the final sample, the sources with inconsistent OPAs in the two surveys (unsuited sources in text) in the LS-SDSS sample are excluded in the analyses.}
    \label{fig:OPAs}
\end{figure}

To obtain reliable jet-galaxy alignment measurements, it is also important to have reliable OPA measurements.
Considering that many sources in the Legacy Surveys also have detections in the SDSS and the PSF sizes in the two surveys are comparable \citep{Dey19}, we assessed the reliability of OPAs in the Legacy Surveys by comparing the OPA measurements in the two surveys, because neither the Legacy Surveys nor the SDSS provided errors for the OPAs.
Because the OPAs in the two surveys were measured independently, sources with OPA that can be measured reliably should have consistent OPAs in the two surveys.\footnote{The shape parameters in the Legacy Surveys are fitted based on the combined images from all three bands ({\it grz}), therefore the morphologies are the same \citep[see Sect. 8 in][]{Dey19}. The OPAs from the SDSS are based on the `{\tt deVPhi\_r}' parameter in the database. We have confirmed that the OPAs in the SDSS are consistent in different bands for our sample.}
We show the OPAs of sources with a counterpart in both the Legacy Surveys and the SDSS (`LS-SDSS sources' hereafter) in our sample in Fig. \ref{fig:OPAs}.
While most of these sources have consistent OPAs in the two optical surveys, a large number of sources have a poorly-defined OPA$_{\rm LS}$ (0\degree in the Legacy Surveys) or inconsistent OPA.
\begin{figure*}
    \centering
    \includegraphics[width=\linewidth]{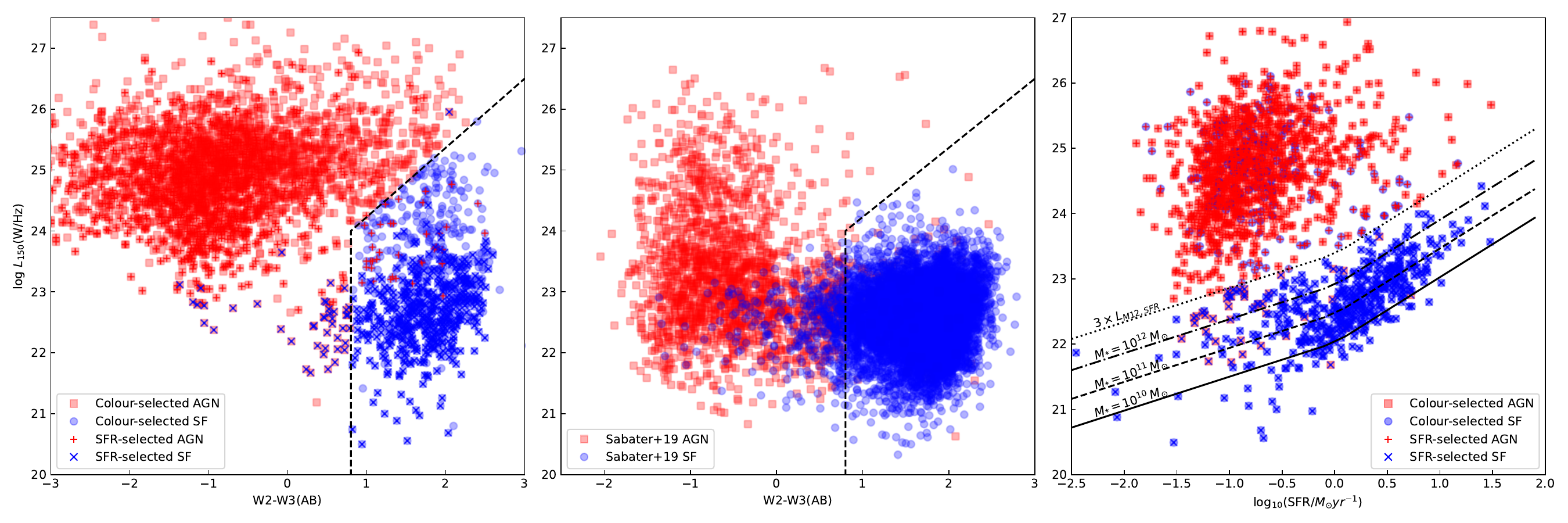}
    \caption{Radio AGN/SF galaxies classification diagrams.
    %
    Left: the radio luminosity and W2-W3 AB colours for the radio sources in the LowzRO sample in this work.
    The red squares and blue circles show the radio AGNs and SF galaxies classified based on the WISE colour respectively.
    The red `+' and blue `x' markers show the radio AGNs and SF galaxies classified based on the SFR from the MPA-JHU value-added catalogue \citep{Brinchmann04}.
    Middle: the radio luminosity and W2-W3 AB colours for the sources in \citet{Sabater19} based on the low redshift LoTSS DR1 sample ($z<0.3$).
    The red squares and blue circles denote the radio AGN and SF galaxies classified based on radio luminosity, spectroscopic information and WISE colours in \citet{Sabater19}.
    The dashed lines in the left and middle panels show the AGN/SF division lines used in the WISE-colour classifications.
    Right: the radio luminosity and the SFR for the radio sources in the LowzRO sample with SFR measurements.
    The symbols have similar meaning as in the left panel.
    The black solid line shows the expected \lrad from star formation for galaxies with \mstar=$10^{10}\,M_{\odot}$ following the relation in \citet{Gurkan18}.
    The black dashed, dot dashed lines show the expected \lrad from star formation for galaxies with \mstar=$10^{11}\,M_{\odot}$ and $10^{12}\,M_{\odot}$ respectively.
    The dotted line shows the division line used in the SFR-base classification in this work, i.e. three times the expected \lrad for galaxies with \mstar=$10^{12}\,M_{\odot}$.
    }
    \label{fig:AGN}
\end{figure*}
For each LS-SDSS source, we obtained the OPA difference between these two surveys, $\Delta$OPA=$\rm |OPA_{LS}-OPA_{SDSS}|$.
We investigated the change of the $\Delta$OPA as a function of different morphological parameters in the Legacy Surveys and calculated the fraction of `unsuited sources' ($\Delta$OPA$>20$\degree).
We found that the unsuited sources were primarily sources with a size close to the PSF size or with a round shape (large axis ratio).
Therefore, to obtain a sample with reliable OPA measurements, we include three more criteria in the sample selection process:
\begin{itemize}
    \item[(6)] half-light radius in the Legacy Surveys $R_{50}>1.5$\arcsec;
    \item[(7)] $R_{50}/{\rm FWHM}_{\rm PSF}>1.2$, where ${\rm FWHM}_{\rm PSF}$ is the FWHM of the PSF in the Legacy Surveys;
    \item[(8)] optical axis ratio $b/a$<0.8.
\end{itemize}
After adopting these criteria, we show the OPAs of the remaining LS-SDSS sources in Fig. \ref{fig:OPAs}.
We found that the sources with poorly-defined OPA$_{\rm LS}$ or inconsistent OPA were mostly removed.
The fraction of unsuited sources in the remaining LS-SDSS sample was $\lesssim10\%$. 
These unsuited sources identified in the LS-SDSS sample are also excluded in the final sample. 

We also only consider sources with $z<0.5$ to avoid possible evolutionary effects.
This also excludes distant barely resolved sources.
With this redshift cut, we selected 4537 radio sources with reliable RPA and OPA in the LoTSS DR2 VAC and the Legacy Surveys.
For simplicity, we call this sample the `LowzRO' sample hereafter.
In the LowzRO sample, 3257 sources have spectroscopic redshifts (3220 from the SDSS DR16, 36 from the DESI spectroscopic survey and one from the HETDEX), the other 1280 sources have a photometric redshift from \citet{Duncan22}.  

\subsubsection{Radio AGN classification}\label{sec:agnsf}
The radio emission from galaxies can be due to AGN activities or star formation \citep[e.g.][]{Condon92,Best05a,Sabater19}.
The radio emission from the AGNs usually takes the form of collimated jets, the direction of which should be related to the spin of the SMBH.
In contrast, the emission from star formation would be spread across a galaxy.
Therefore, the radio emission of the SF galaxies is expected to be aligned with the optical major axis of the galaxy.
This would result in a contamination in the jet-galaxy alignment analyses.
To reduce the contamination from the SF galaxies, we need to identify the radio AGNs in the LowzRO sample.

To separate the radio AGNs and star-forming galaxies, we first used the WISE colours because all sources in the LowzRO sample have WISE measurements.
As the W2 band captures the stellar emission and the W3 band the Polycyclic Aromatic Hydrocarbon (PAH) strength \citep{Jarrett11}, which is an indicator for warm dust associated with SF, the W2-W3 colour is used to distinguish between star-forming and passive galaxies \citep{Yan13,Cluver14,Herpich16}.
Because the host galaxies of radio AGNs and SF galaxies tend to have different star formation rate \citep[SFR; e.g.][]{Heckman14}, the W2-W3 colour has also been used to select radio AGN in \citet{Sabater19} and \citet{Hardcastle19}.
Only using the W2-W3 colour might only recover part of the radio AGNs population as it does not take the radio information into account explicitly.
We therefore introduce the radio luminosity-WISE colour diagram to the radio AGN selection process.

We show the logarithmic radio luminosity at 150 MHz log($L_{150\,\rm MHz}/(\rm W\,Hz^{-1})$) as a function of the W2-W3 colour for the low redshift sources ($z<0.5$) in the LowzRO sample in the left panel of Fig.\ref{fig:AGN}.
The WISE colours are calculated in AB magnitudes for comparison with the results in \citet{Sabater19}.
It shows that the sources can be divided into two populations, which cluster in two regions in the radio luminosity-WISE colour diagram.
As suggested by \citet{Gurkan18}, since the radio emission from star-formation is unlikely to reach $L_{150\,\rm MHz}>10^{25}\rm\, W\,Hz^{-1}$, the population in the top left region in the radio luminosity-WISE colour diagram (high radio luminosity, small W2-W3) should be associated with radio AGNs.
The bluer and less luminous population should be SF galaxies.

For comparison, we also plot the radio sources in \citet{Sabater19} with detailed AGN/SF classifications based on the LoTSS DR1 data with a smaller sample size than the DR2 in the right panel of Fig.\ref{fig:AGN}.
While the radio sources in \citet{Sabater19} in general have lower radio luminosity, they also show a similar bimodality in the radio luminosity-WISE colour diagram with most of their SF galaxies in the bluer region and the radio AGNs in the red region.
This is not surprising because \citet{Sabater19} used the W2-W3 colour as an auxiliary method in their AGN/SF classification.
They used a simple division of W2-W3=0.8 and found the WISE-colour classifications were highly consistent with the classifications based on the radio luminosities and spectroscopic information.
But this simple division would fail to separate the two populations for high luminosity sources in the LowzRO sample.

To make our classification consistent with previous results, we classified low luminosity sources ($L_{150\rm\,MHz}<10^{24}\,\rm W\,Hz^{-1}$) with W2-W3<0.8 as radio AGNs, while at high luminosities we chose radio AGNs to be the sources with ${\rm log}(L_{150\,\rm MHz}/(\rm W\,Hz^{-1}))>24+(W2-W3-0.8)\times2.08$, which was approximated based on human-eye inspection on the \lrad-WISE colour diagram to separate the two clusters well.
We applied our division lines for the sample in \citet{Sabater19} in the middle panel of Fig. \ref{fig:AGN} and they were consistent with the AGN/SF classifications at high luminosities. 
Our AGN/SF classifications successfully reproduced $\sim$95\% of the classification results in \citet{Sabater19}, which were based on more sophisticated methods.

The AGN/SF classifications can be obtained by estimating the radio emission contributed from the SFR.
In the LowzRO sample, 1696 sources have SFR measurements in the Max Planck for Astrophysics and Johns Hopkins University groups’ (MPA-JHU) value-added catalogue \citep{Brinchmann04}. 
We plot the radio luminosity of these sources as a function of their SFR in the right panel of Fig. \ref{fig:AGN}.
We also showed the \lrad-SFR relations for SF galaxies with different stellar mass from \citep{Gurkan18} in Fig. \ref{fig:AGN}.
Apparently, our colour-selected AGNs and SF galaxies are well separated and the AGNs have much higher \lrad than the expected \lrad based on SFR.
This also supports our AGN/SF classifications based on the WISE colours.

For sources with SFR measurements, we select sources with a \lrad\, larger than three times the expected SFR-based \lrad\, for a \mstar=$10^{12}\,M_{\odot}$ galaxy to be SFR-selected AGNs and the others to be SFR-selected SF galaxies.
Although most of the sources in our sample are less massive than $10^{12}\,M_{\odot}$, we choose a \mstar\, of $10^{12}\,M_{\odot}$ in the AGN classifications to have a conservative result, which could reduce the rate of false positives. 
As a result, 987(44) out of 1031 colour-selected AGNs are also SFR-selected AGNs (SF galaxies), and 155 (433) out of 588 colour-selected SFs are SFR-selected AGNs (SF galaxies).

Moreover, we matched the LowzRO sample with the giant radio galaxy (GRG) catalogue from \citet{Oei23}.
These GRG have very extended radio structures ($\gtrsim$0.7 Mpc) produced by powerful radio jets, therefore they should also be classified as radio AGNs.
In this way, 21 more sources are added to the AGN sample.

To combine these classification methods, we adopt the classification based on the SFR when the SFR measurement is available.
Based on the three methods, we classified 3121 sources as low redshift radio AGNs and 1395 as low redshift SF galaxies (the source numbers are listed in Tab. \ref{tab:samp}).
We call them `LowzROAGN' and `LowzROSF' in the following analyses.
While our W2-W3 division lines showed a classification accuracy of about 95\% for the low luminosity sample in \cite{Sabater19}, we suggest that our classifications would have a higher accuracy in our sample, which have a larger fraction of radio sources with \lrad$>10^{24}\,\rm W\,Hz^{-1}$.

\subsubsection{FIRST sample}
To compare with previous work \citep{2009MNRAS.399.1888B}, we also used data from the FIRST survey.
The FIRST survey covers 10,000 square degrees of the north and south galactic caps \citep{Becker95} and produced roughly 946,000 sources at a resolution of $\rm{5\arcsec}$ and a typical rms of 0.15 mJy beam$^{-1}$ over 10575 square degrees \citep{Helfand15}. At the detection threshold of $\rm{1\,mJy}$, the number of sources per square degree is roughly 90, and approximately 35\% of the sources have resolved structure on $\rm{2}$-$\rm{30\arcsec}$ scales with 30\% of sources having counterparts in SDSS \citep{2022MNRAS.509.4024D}.
We also use data from the National Radio Astronomy Observatory (NRAO) VLA Sky Survey \citep[NVSS;][]{Condon98} for additional information in the sample selection process.
The NVSS catalogue covers the whole sky area north of Declination -40\degree with an Stokes I rms of 0.45 mJy beam$^{-1}$ at 1.4 GHz and a resolution of about 45\arcsec .
We used total flux densities from NVSS rather than FIRST for sources with NVSS detections since NVSS is more sensitive to extended emission.

We used the radio sample described in \citet{Best12}, which contains 18\,286 radio sources with an optical counterpart in the SDSS data release 7 \citep{Abazajian09}. 
These sources were further cross-matched with the photometric and spectroscopic data using {\tt CasJobs}\footnote{\url{https://skyserver.sdss.org/CasJobs/}} based on their object IDs. 
For these sources, we obtained the OPAs and host galaxy properties including stellar mass ($M_{\star}$), SFR, axis ratio ($b/a$; $a$ and $b$ are apparent major and minor axes of the host galaxy), the flux of the doubly ionized oxygen ([O[III]) emission line, and the velocity dispersion (${\sigma}$). 
The host galaxy properties were based on the MPA-JHU spectroscopic reanalysis within the SDSS database \citep{Brinchmann04}. 
In this way, we obtain 17,165 sources. 
We also applied the restrictions mentioned in \citet{2009MNRAS.399.1888B}.
We excluded faint galaxies with $r>18$ where $r$ is the $r$ band magnitude, round galaxies with the axis ratio larger than 0.8 in either optical or radio band and all radio sources with a deconvolved angular size smaller than 2\arcsec in the FIRST survey.
To ensure a reliable RPA measurement, we further require the radio sources to either have a large angular size or have consistent RPA measurements in both the FIRST survey and the NVSS.
We cross-matched the radio sources with the NVSS catalogue with a 30\arcsec\, searching radius.
For radio sources with a deconvolved angular size smaller than 8\arcsec\, in the FIRST catalogue, we compared their RPAs in the FIRST and the NVSS catalogue and excluded all sources with an RPA difference larger than 20\degree .
We adopted the RPAs measurement from the FIRST survey for radio sources with a deconvolved size larger than 8\arcsec\, in the FIRST catalogue regardless of their RPAs in the NVSS, because the smaller scale structures are more related to the spin direction of SMBHs. 
We note that this constraint on the size of radio sources is more conservative than that used by \citet{2009MNRAS.399.1888B}. 
This 8\arcsec\, threshold is the same as the one used for the LoTSS sources because the two radio surveys have similar resolutions.
We also require our sources to be large sources with $r$ band half-light radius $R_{50}>1.5\arcsec$.
In addition, we only took radio AGNs classified in \citet{Best12} using methods similar to  \citet{Sabater19}. 

After the restrictions, we obtained a sample called `FIRSTAGN' containing 1\,035 AGNs (Tab. \ref{tab:samp}) with redshift $z<0.5$ ($\sim 95\%$ have $z<0.3$), and they have better data quality as those in \citet{2009MNRAS.399.1888B}.
The RPA distribution of the selected sources does not show significant bias and \citet{2009MNRAS.399.1888B} also suggested that the OPAs are reliable.
In the FIRSTAGN and the LowzROAGN sample, we found that 177 pairs of sources have the same SDSS counterparts, and another 5 pairs of sources are within a separation of 5\arcsec\, and have similar OPAs, RPAs and b/a.
These overlapping sources have independent but consistent RPA and OPA measurements in the two samples.
We therefore conclude that the RPAs and OPAs in the FIRSTAGN sample are reliable and consistent with those in the LowzROAGN sample. 
\begin{figure}
    \centering
    \includegraphics[width=\linewidth]{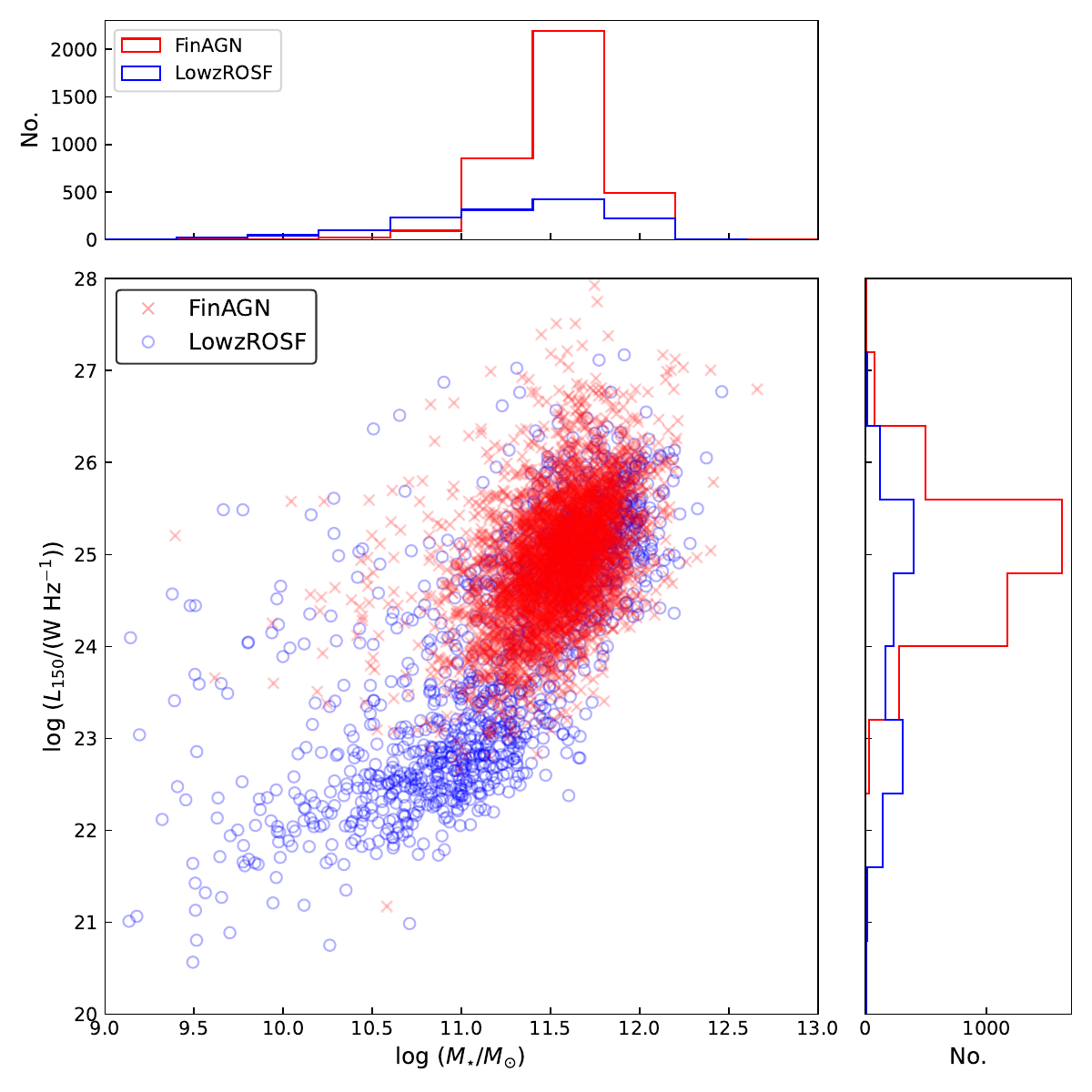}
    \caption{The $M_{\star}$ and $L_{150\rm\,MHz}$ distributions of the radio samples selected in this work.
    In the main panel, sources in the FinAGN and LowzROSF samples are denoted by red crosses and blue empty circles respectively.
    It should be noted that many blue circles are covered by the red crosses because of the overlapping parametric space.
    Top: the $M_{\star}$ distributions of the four samples. 
    The FinAGN and LowzROSF samples are shown as red and blue histograms respectively.
    Right: the $L_{150\rm\,MHz}$ distributions of the two samples.
    }\label{fig:dist}
\end{figure}

\begin{figure}
\centering
\includegraphics[width=\linewidth]{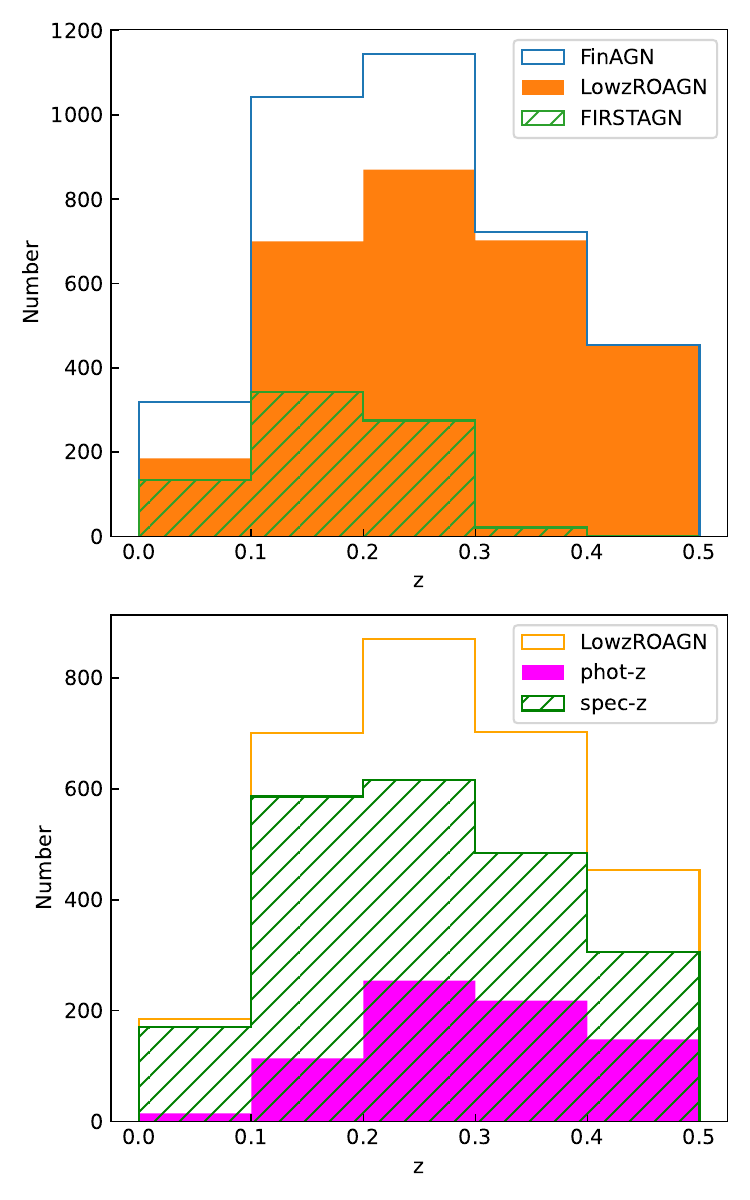}
\caption{
Top: the redshift distributions of the radio AGNs selected in this work.
The blue, orange (filled) and green (hatched) histograms show the distributions for the FinAGN, LowzROAGN and FIRSTAGN sample (after visual check) in Tab. \ref{tab:samp} respectively.
Bottom: the redshift distributions of the LowzROAGN sample.
The orange, magenta (filled) and green (hatched) histograms show the distributions for the whole LowzROAGN sample, the sources with photometric redshifts and spectroscopic redshifts respectively.
}\label{fig:zdist}
\end{figure}

\begin{table*}
    \centering
    \begin{tabular}{c|ccc}
        Sample & Number of sources & Removed & In final sample\\
        \hline
        LowzROAGN & 3\,121 & 241& 2880\\
        FIRSTAGN & 1\,035 & 233$^c$ & 802\\
        LowzROSF$^a$ & 1\,395 & - & - \\
        FinAGN$^b$ & 3\,682 & - & - \\
    \end{tabular}
    \caption{The number of sources in the low redshift radio samples selected in this work. ($a$): We did not perform visual checks of LowzROSF because it is not our main topic in this work, therefore only the total source number is listed here. ($b$): FinAGN is the sample after all the removal processes. We only listed the final source number here. ($c$): Duplicated sources are also removed from the FIRSTAGN sample.}\label{tab:samp}
\end{table*}

\subsection{Additional cleaning of the sample}
Previous automated processes have reduced our radio AGN sample size to a few thousands.
However, there could still be some sources with incorrect RPA and OPA measurements due to mergers, low brightness or incorrect radio-optical associations.
To further improve the reliability of the RPA and OPA measurements of our sample, we selected and checked some sources with visual inspection.

First, we selected all sources with a host galaxy having a neighbor within 6\arcsec (approximately the beam size of the LoTSS and the FIRST surveys) in the DESI Legacy Surveys. 
These sources could have inaccurate OPAs because of mergers or interaction features, or incorrect radio-optical associations.
Second, we chose the sources in the LowzROAGN sample with $r$ band magnitude fainter than 19\footnote{The DESI Legacy Surveys have typically better sensitivity than the SDSS \citep{Dey19}, therefore we chose a threshold one magnitude fainter than that the threshold for the FIRSTAGN sample.} to check the reliability of their OPA measurements.
Third, due to the rareness of disky galaxies with powerful jets \citep{Barisic19,Zheng20}, all sources with b/a less than 0.4 are also selected for visual check.
As a result, we selected 325 sources from the FIRSTAGN sample and 1312 sources from the LowzROAGN sample for further visual inspection.

During our inspection of radio and optical images we verified the RPA, OPA, and other shape parameters of objects listed in the catalogue.
For sources from the LowzROAGN sample, we carefully checked the radio image from the LoTSS and the optical images in $r$, $g$, $z$ bands from the DESI Legacy Surveys.
For sources from the FIRSTAGN sample, we checked their radio images from both the FIRST and the NVSS, and the SDSS optical images in $r$ band.   

Sources with at least one of the following problems are flagged and excluded from the final sample:
\begin{itemize}
    \item The determination of the source morphology is clearly affected by neighboring sources and the OPA measurement would change significantly if there was no neighboring sources.
    \item The source has an irregular shape or signs of two distinct cores.
    \item The radio emission has a complex structure and a reliable RPA could not be determined.
    \item Two or more optical sources could be associated with a radio source.
\end{itemize}
This process removed 50 sources from the FIRSTAGN and 226 sources from the LowzROAGN sample. 

Furthermore, there are 12 sources classified as narrow angle tailed sources (NATs) in \citet{Sasmal22}. 
These sources are also excluded from the final sample since they could have ill-defined RPAs. 
We also noticed some peculiar sources that we removed from the final sample, which are listed as follows.
\begin{itemize}
    \item {\it ILT J105831.54+564347.0}: A large NAT reported in \citet{Wilber18}.
    \item {\it ILT J125455.67+270936.9}: NGC 4789, a large bent-tailed radio source \citep[see e.g.][]{Venturi89,Bonafede22}.
    \item {\it ILT J094043.09+305914.1}: The optical image shows a close pair and the true host of the radio source is not certain.
    \item {\it ILT J154912.62+304715.5}: A famous Einstein ring MG 1549+305 \citep{Lehar93,Treu03}. The host of the radio source is too faint to assign an OPA.
\end{itemize}

\subsection{Summary of the samples}
In summary, we selected two low redshift radio AGN samples and an SF galaxy sample based on the LoTSS DR2 and the FIRST data.
We list the number of sources in these samples in Tab. \ref{tab:samp}.
The LowzROAGN and LowzROSF are the low-redshift ($z<0.5$) radio AGNs and SF galaxies with reliable PAs in both radio and optical bands selected from the LoTSS DR2, and the FIRSTAGN are radio AGNs selected from \citet{Best12}.
For the following analyses, we merged the FIRSTAGN and LowzROAGN samples into a final radio AGN sample `FinAGN' containing 3682 sources.
For the sources appearing in both samples, their RPA and OPA are determined by the measurements in the LowzROAGN sample.

For the radio samples based on the LoTSS DR2, we estimated their stellar masses using the $M_{\star}-\nu L_{\nu}(3.4\,\rm \mu m)$ relations \citep{Wen13} for AGNs,
\begin{equation}
    {\rm log}(\frac{M_{\star}}{M_{\odot}}) = 1.132+{\rm log}(\frac{\nu L_{\nu}(3.4\,\rm \mu m)}{L_{\odot}}) 
\end{equation}
and for SF galaxies,
\begin{equation}
    {\rm log}(\frac{M_{\star}}{M_{\odot}}) = -0.04+1.12\,{\rm log}(\frac{\nu L_{\nu}(3.4\,\rm \mu m)}{L_{\odot}})
\end{equation}
where $\nu L_{\nu}(3.4\,\rm \mu m)$ is the rest frame 3.4\,$\mu$m radio luminosity.
The derived \mstar are consistent with the results based on spectroscopic analysis for the sources with SDSS data with a difference of $\lesssim$0.2 dex.
This was also confirmed in \citet{Wen13}.

We show the $M_{\star}$ and $L_{150\rm\,MHz}$ distributions of the FinAGN and LowzROSF samples in Fig. \ref{fig:dist}.
For sources from the FIRSTAGN sample, the \lrad are calculated assuming a canonical radio spectral index of 0.7 \citep{Condon02} based on the 1.4 GHz flux density from the NVSS catalogue.
From Fig. \ref{fig:dist}, we can see that our radio AGN samples typically have $M_{\star}$ ranging from $10^{11}$ to $10^{12}\,M_{\odot}$ and $L_{150\rm\,MHz}$ ranging from $10^{23}$ to $10^{27}\,\rm W\,Hz^{-1}$.
In contrast, the SF galaxies have a higher fraction of low $M_{\star}$ and low $L_{150\rm\,MHz}$ sources.
The distributions of the redshifts for sources in the FinAGN sample are shown in Fig. \ref{fig:zdist}. 

We also note that some sources that do not have a reliable deconvolved minor axis (`Min<6' in the LoTSS DR2 VAC) could have a reliable large major axis (`Maj>10').
These radio sources might also have a well determined RPA.
However, only about 20 sources meet the criteria for AGNs described in Sect. \ref{sec:rpa}-\ref{sec:agnsf}. 
This small number would not impact the following statistical analyses.
Therefore, for simplicity and consistency, we do not include these sources in the final sample.

    \begin{figure}
    \centering
    \includegraphics[width=\linewidth]{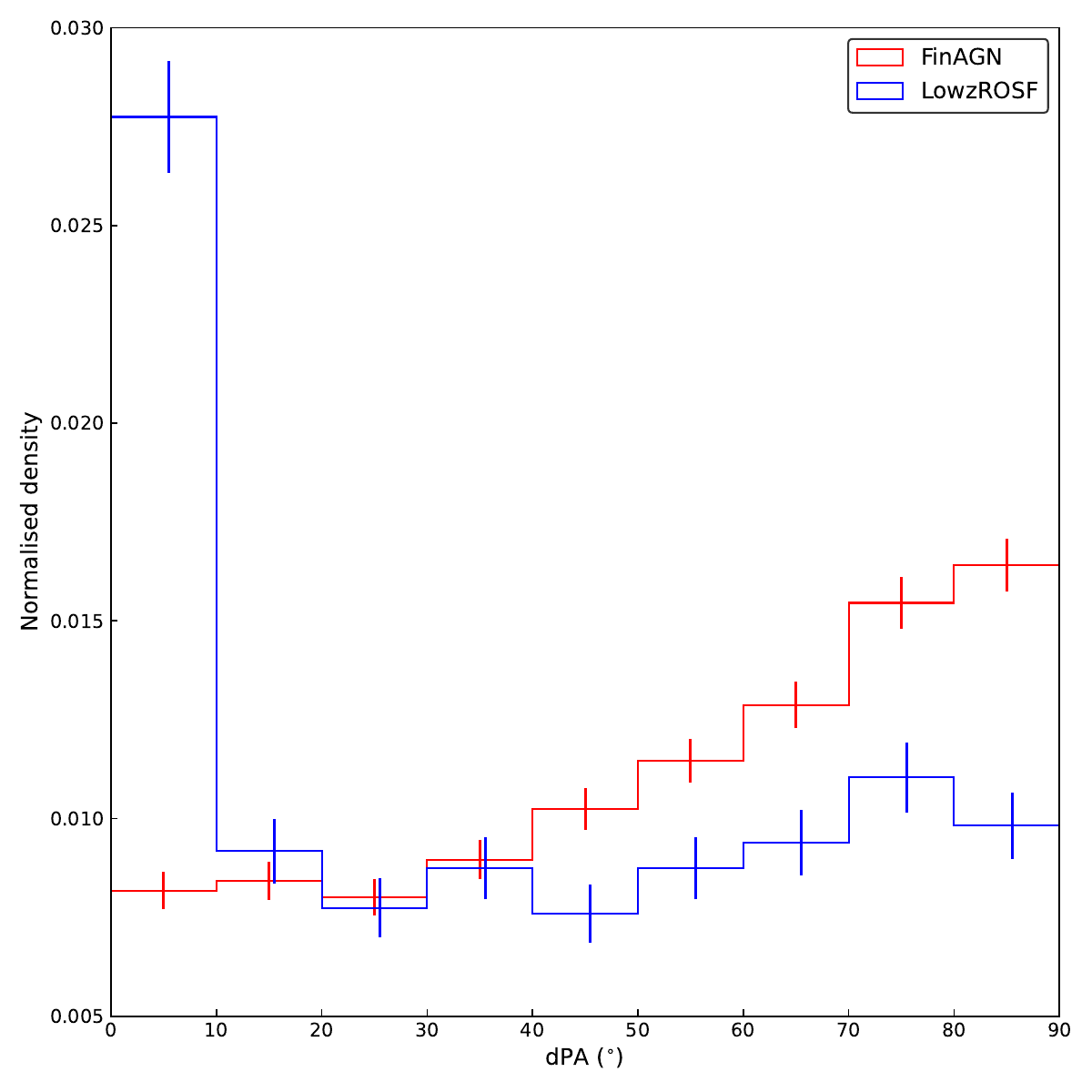}
    \caption{The alignment distributions for the two samples.
    %
    The FinAGN and LowzROSF samples are shown as red and blue histograms respectively.
    The error bars for the LowzROSF sample are offset for clarity.
    The histograms are normalised to make the integration of the histograms equal to one.
    The errorbars (and the errorbars in the histograms throughout the work) are derived from Poissonian errors.}
    \label{fig:2}       
    \end{figure}

\section{Radio-optical alignment distribution}
\label{roalign}

In the following analyses, we simply assume that the radio emission from the FinAGN sample is produced by radio jets, because these sources all have extended structure and over 99\% of them have a high radio luminosity (\lrad$>10^{23}\rm\,W\,Hz^{-1}$).   
To investigate the orientation of the radio jets with respect to their host galaxies, we defined the radio-optical misalignment angle of a galaxy, labelled as dPA, to be the apparent difference between the galaxy's RPA and OPA.  
In Fig. \ref{fig:2}, we show the dPA distributions for the FinAGN and LowzROSF samples.

It is apparent that the dPA of the radio AGNs and SF galaxies are significantly different. 
The distributions of dPA for the radio AGNs are biased towards 90\degree. 
This prominent minor-axis alignment tendency indicates that the apparent jet orientations are more likely to be perpendicular to their optical major axis.
The fraction of AGNs with dPA$>60$\degree is 1.8 times of the fraction of the AGNs with dPA$\leq30$\degree. 
In contrast, the dPAs of the SF galaxies are significantly biased towards 0\degree, indicating that the radio emission of the galaxies is aligned with the optical major axis.
This major-axis alignment tendency agrees with the idea that the radio emission from the SF galaxies originates from the star formation activity across the galactic plane. 
We noticed that a small fraction of the SF galaxies also have a large dPA.
This could be due to the contribution from AGNs because we have used a conservative classification in the last section to reduce the rate of false positive in the AGN sample, which might rule out some low luminosity AGNs.
Since we mainly discuss the properties of radio AGNs in this work, this contamination is not relevant in this work.

    \begin{figure*}
    \centering
    \includegraphics[width=\linewidth]{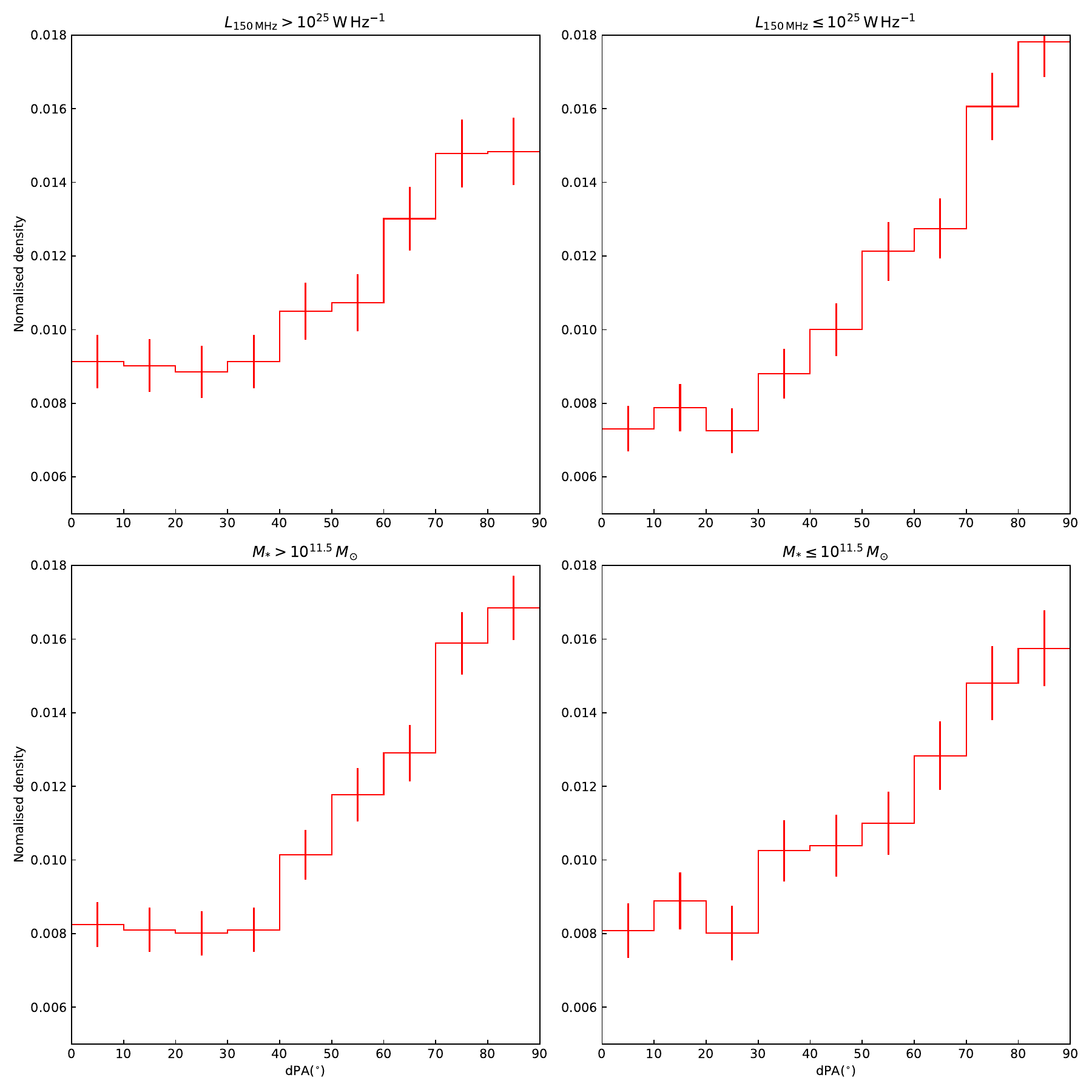}
    \caption{Radio-optical misalignment angle distributions for radio AGNs with different $L_{150\,\rm MHz}$ and \mstar. 
    Top (left to right): The radio-optical misalignment angle distributions of moreL ($L_{150\,\rm MHz}>10^{25}\,\rm W\,Hz^{-1}$) and lessL ($L_{150\,\rm MHz}\leq10^{25}\,\rm W\,Hz^{-1}$) radio AGNs.
    Bottom (left to right): The radio-optical misalignment angle distributions of massive (\mstar$>10^{11.5}\,M_{\odot}$) and less massive (\mstar$\leq10^{11.5}\,M_{\odot}$) radio AGNs.
    }
    \label{fig:3}       
    \end{figure*}
    
    \begin{figure*}
    \centering
    \includegraphics[width=\linewidth]{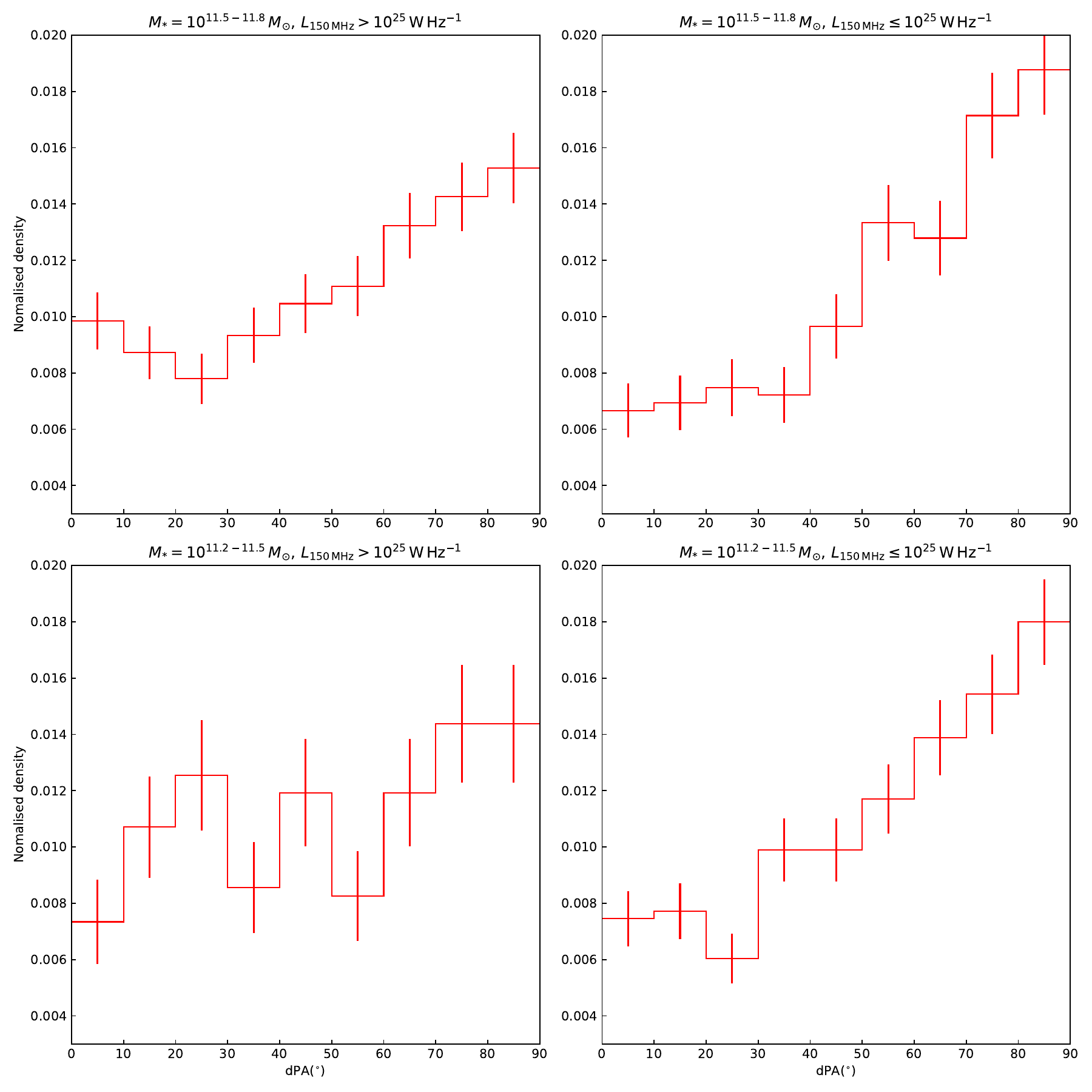}
    \caption{Radio-optical misalignment angle distributions of radio AGN with different radio luminosities in two \mstar bins. 
    Top (left to right): The radio-optical misalignment angle distributions of moreL ($L_{150\,\rm MHz}>10^{25}\,\rm W\,Hz^{-1}$) and lessL($L_{150\,\rm MHz}\leq10^{25}\,\rm W\,Hz^{-1}$) radio AGNs with \mstar ranging from $10^{11.5}$ to $10^{11.8}$ $M_{\odot}$.
    Bottom (left to right): The radio-optical misalignment angle distributions of moreL and lessL radio AGNs with \mstar ranging from $10^{11.2}$ to $10^{11.8}$ $M_{\odot}$.
    }
    \label{fig:4}       
    \end{figure*}
\subsection{$L_{150\,\rm MHz}$ and $M_{\star}$ divisions}\label{sec:LM}
Radio AGNs with different luminosities and \mstar\, exhibit different physical properties \citep[see e.g.][]{Janssen12,Heckman14,Zheng20}, which implies different physical processes in these AGNs.
In this section, we further investigate the dPA for sources with different \mstar\, and radio luminosity, by dividing the samples into different groups for alignment analysis based on their radio luminosity $L_{150\,\rm MHz}$ and the $M_{\star}$ of their host galaxy.

Based on the $L_{150\,\rm MHz}$ distributions, to ensure comparable numbers of sources in different groups, we divided the entire sample into higher and lower radio luminosity sources depending on whether they have $L_{150\,\rm MHz}$ larger or smaller than $10^{25}\,\rm W\,Hz^{-1}$, labelled as `moreL' and `lessL' for simplicity hereafter. 
The two groups contain 1752 and 1930 sources respectively.
The corresponding dPA distributions are shown in the upper panels of Fig. \ref{fig:3}. 
The radio AGNs are generally minor-axis aligned regardless of their luminosities, while the dPA distribution seems to flatten at both ends of the dPA range for the moreL sample.
We performed a Kolmogorov-Smirnov (K-S) test on the two dPA distributions and the result showed a null hypothesis probability ($p_{\rm null}$) of 0.5\% .
This suggests that the two dPA distributions are not drawn from the same population.
The moreL radio AGNs tend to have less minor-axis aligned sources.

\begin{figure*}
    \centering
    \includegraphics[width=\linewidth]{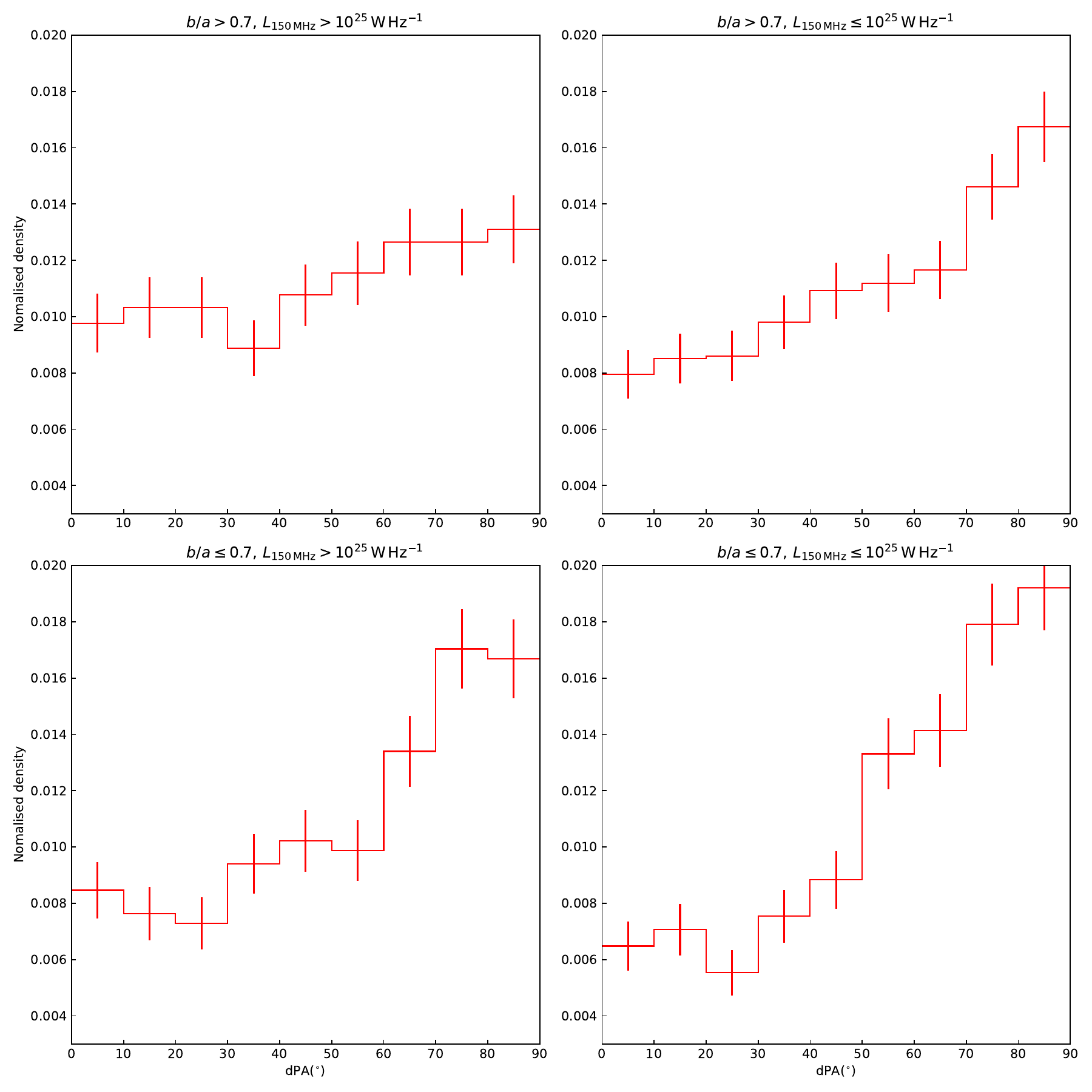}
    \caption{Radio-optical misalignment angle distributions of radio AGN with different radio luminosities in two groups with different axis ratio. 
    Top (left to right): The radio-optical misalignment angle distributions of moreL and lessL radio AGNs with $b/a>0.7$.
    Bottom (left to right): The radio-optical misalignment angle distributions of moreL and lessL radio AGNs with $b/a\leq0.7$.
    }
    \label{fig:ba}    
\end{figure*}

\begin{figure*}
    \centering
    \includegraphics[width=\linewidth]{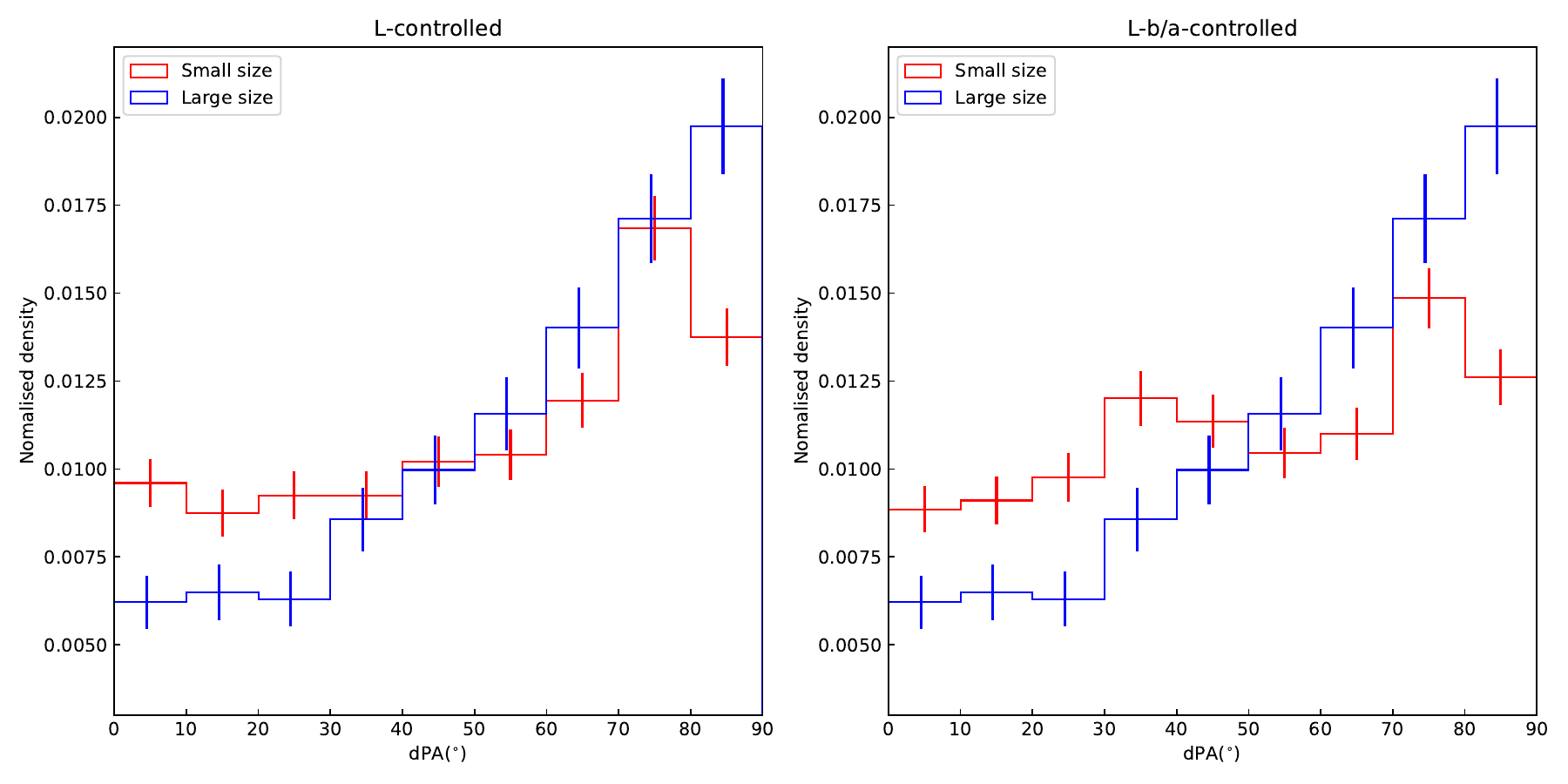}
    \caption{Radio-optical misalignment angle distributions of radio AGN with different projected linear size in samples with similar \lrad distribution or similar \lrad-$b/a$ distribution.
    Left: The dPA distributions of radio AGNs with different projected linear size and similar \lrad distribution.
    Right: The dPA distributions of radio AGNs with different projected linear size and similar \lrad-$b/a$ distribution.
    In both panels, the blue histograms denote radio AGNs with large projected linear size ($>250$\,kpc) and the red histograms denote radio AGNs with small size ($\leq250$\,kpc).
    The errorbars are slightly offset for clarity.}\label{fig:size}
\end{figure*}
The sample was also separated into massive and less massive groups based on whether their \mstar\, is larger than $10^{11.5}\,M_{\odot}$.
The two groups contain 2209 and 1473 sources respectively.
The dPA distributions of massive and less massive galaxies are shown in the lower panels of Fig. \ref{fig:3}.
Similar to the \lrad-divided results, the dPA distributions show similar minor-axis alignment tendencies in both two groups, but the less massive galaxies have a lower peak while the massive galaxies show a slight up-turn below 30\degree.
The K-S test result indicates a $p_{\rm null}$=21.3\%, which cannot rejects the null hypothesis that the two dPA distributions are drawn from the same population.
Therefore the dPA distribution is not likely to be \mstar\, dependent.
    
The up-turn tendency at small dPAs in the dPA distribution is likely to be related to both \lrad\, and \mstar, hence purely \lrad\, or \mstar-divided analyses might weaken the trend. 
To further investigate this weak major-axis alignment tendency and check the influence of \lrad\, and \mstar\, on the dPA distribution, we check the dPA distributions in two narrow \mstar\, bins with large numbers of sources. 
The first category includes AGNs with \mstar\, between $10^{11.5}$ and $10^{11.8}\,M_{\odot}$, while the second category includes AGNs with \mstar\, between $10^{11.2}$ and $10^{11.5}\,M_{\odot}$. 
These two categories of AGNs were then further divided according to whether they are moreL or lessL sources. 
    
Fig. \ref{fig:4} shows the dPA distributions of radio AGNs with different \lrad in the two \mstar\, bins.
The numbers of sources in the four groups are 975, 735, 327 and 778 respectively (from top left to bottom right in Fig. \ref{fig:4}).
Generally, the dPA distributions in the four groups all show a minor-axis alignment.
We performed K-S tests to compare the dPA distributions in every two groups with similar \mstar\, or similar \lrad.
For groups with similar \mstar, both of the two the K-S test results indicate $p_{\rm null}$ smaller than 5\% (0.2\% and 1.5\%), while the results for groups with similar \lrad have $p_{\rm null}$ much larger than 5\%.
Therefore, for samples with similar \lrad, the K-S test cannot reject the null hypothesis that they are drawn from the same population. 
These results further show that the dPA distributions are more likely to be \lrad dependent rather than \mstar\, dependent.
It is also clear from the histograms that sources with higher luminosity are less likely to be minor-axis aligned.

The stronger major-axis alignment tendency in the moreL\&massive samples in the top left panel of Fig. \ref{fig:4} implies that the major-axis alignment tendency is not likely to be the result of the contamination by SF galaxies, because the SF contamination should be largest for the less massive and low-luminosity populations \citep{Best12,Sabater19}.
A similar trend was suggested in \citet{Andernach09} for the brightest cluster members.
However, we checked the images of AGNs with dPA$<10$\degree in the moreL\&massive sample and cannot find any special features related to environment or morphology compared to other AGNs.
Given that the weak major-axis alignment peak in AGNs with $\rm dPA<10$\degree is due to only 10--20\% of
all sources below dPA=10\degree (see Sect. \ref{sec:intrinsic}), it is difficult to identify the sources causing the excess in our statistical survey.
Deeper multiwavelength imaging should help in verifying these special sources in the future.

\subsection{Galaxy axis ratio dependence}\label{sec:Lmor}
Recent studies suggested that powerful radio jets tend to reside in round galaxies \citep[e.g.][]{Barisic19,Zheng20}.
This trend may be the result of the central black holes of galaxies with a merger-dominant assembly history having a different distribution of spin compared to galaxies with a more quiescent history \citep{Zheng23}. 
If this hypothesis is correct, then we should be able to see different dPA distributions for round and elongated galaxies.

We chose the axis ratio $b/a=0.7$ to divide the samples into two groups with comparable numbers of sources.
The dPA distributions of the radio AGNs with different \lrad\, in the two groups are shown in Fig. \ref{fig:ba}.
The four groups contain 901, 1081, 851 and 849 sources for the moreL and lessL sources with $b/a>0.7$, and for the moreL and lessL sources with $b/a\leq0.7$ respectively.
The majority of the dPA distributions still show an overall minor-axis alignment tendency in all four groups.
However, the minor axis alignment tendency seems to be weaker in the two rounder groups. 
For the elongated groups, the number densities of sources with dPA$>70$\degree are about two times the densities at $\leq30$\degree, whereas the highest source density in the moreL\&rounder group are only $\sim30\%$ larger than the lowest point.
This could be a projection effect if the intrinsic jet is roughly aligned with the minor axis of an oblate galaxy. 
In this scenario, as the apparent rounder galaxies are more likely to be face-on, the apparent dPA would become largely affected by the intrinsic offset between the jet and the minor axis of the galaxy and the measurement uncertainties, which tends to flatten the dPA distribution.
This projection effect will be discussed further in Sect. \ref{sec:intrinsic}.

We performed K-S tests on each pair of groups with similar axis ratio ranges or similar \lrad\,(i.e. in the same row or the same column in Fig. \ref{fig:ba}).
For groups with similar \lrad, the K-S tests indicate $p_{\rm null}=0.1\%$ (for moreL groups) and $<0.1\%$ (for lessL groups).
Both of the tests reject the null hypothesis, suggesting significant differences between radio AGNs with different optical axis ratio.
In contrast, for groups with similar axis ratio ranges, the tests result in $p_{\rm null}=2.2\%$ (for rounder groups) and 0.8\% (for elongated groups).
Both tests reject the null hypothesis that the dPA distributions for moreL and lessL radio AGNs are drawn from the same population.
The difference between the dPA distributions of the moreL and lessL samples is still significant.

\subsection{Projected linear size dependence}\label{sec:psize}
In this section, we investigate the correlation between the dPA and the projected linear size of radio sources.
\citet{Palimaka79} claimed that the minor-axis alignment trend is stronger for sources with larger linear size, but it was not confirmed in later works \citep[e.g.][]{VazquezNajar19}.

In LoTSS DR2 VAC, the linear sizes of radio sources were listed as `Size', which were estimated based on redshifts and the largest angular size from manual measurement, `flood-fill size' (described in \citet{Mingo19}), or catalogue-based angular size (`Composite\_Size' or $2\times$`DC\_Maj'\footnote{The deconvolved major axis is the full-width at half-mass of a Gaussian, therefore a factor of 2 is multiplied for the correction to the realistic geometry of the source.}).
Moreover, \citet{Oei23} revisited the size of the large radio sources in the LoTSS DR2 and updated the sizes of GRGs in their works.
We used the sizes from \citet{Oei23} for the GRGs in our sample whenever possible in the following analysis.
It should be noted that the results from \citet{Oei23} were based on radio images with resolutions from 6\arcsec\, to 90\arcsec, which reveals more extended and structure.
Their size definition was not consistent with the other radio sources in the LoTSS DR2 VAC since the diffuse emission would be more important in the measurements for these GRGs.
However, GRGs are rare (only 184) in our sample and possible inconsistencies in size measurements would not influence our statistical results.
%
%
We also found the sizes from the FIRST catalogue could be severely underestimated while the sizes from the NVSS could be inconsistent with the LoTSS measurement because of the much larger resolution.
Therefore, to ensure the consistency and reliability of the size measurements, we dropped the sources without a LoTSS detection in the analysis related to size.

Due to the surface brightness limitation in radio surveys, physically large and low-luminosity source are difficult to detect, as described in \citet{Hardcastle16,Hardcastle19}.
Therefore, large sources are more likely to be high-luminosity sources in our sample.
This will introduce important bias in the dPA analysis because of the difference dPA distributions of moreL and lessL AGNs.

To disentangle the luminosity-size dependence, we construct an L-controlled sample for comparison.
We selected sources with \lrad ranging from $10^{24}$ to $10^{26}\,\rm W\,Hz^{-1}$ in the FinAGN sample with LoTSS detections.
A total of 2453 AGNs are selected and 1221 of them have a projected linear size larger than 250 kpc, while the other 1232 sources have sizes $\leq$250 kpc. 
Based on the \lrad distribution of the AGNs with larger sizes, we resampled 1200 AGNs with smaller sizes to construct an $L$-controlled smaller size AGN sample with the same \lrad distribution as the larger size AGNs.
The dPA of the larger size AGNs and the $L$-controlled smaller size AGNs are shown in Fig. \ref{fig:size}.

Apparently, the larger size AGNs have a stronger minor-axis alignment trend than the L-controlled smaller size sample.
This result is consistent with \citet{Palimaka79}.
It should be noted that in Sect. \ref{sec:LM} the lessL sample have a stronger minor-axis alignment trend than that of the moreL sample.
This is in contrast to the results expected from the dPA distribution of the smaller size AGNs with a weaker minor-axis alignment trend.
Therefore, we suggest that the dPA-\lrad relation is independent to the dPA-size relation.

Since the dPA distribution of radio AGNs suggest intrinsic minor-axis aligned jets, the projection effect could also influence the dPA-size relation.
For an observer with the line-of-sight (LOS) close to the intrinsic minor axis of the galaxy, the projected linear size of the radio jet would be small if the jet is aligned to the minor axis of galaxy.
In this case, the observed dPA distribution will appear less minor-axis aligned.
This is inline with the trend of the dPA-size relation in Fig. \ref{fig:ba} and Fig. \ref{fig:size}.

It is virtually impossible to eliminate the projection effect because the real physical linear sizes of radio sources are not known.
To reduce the projection effect on our results, we use an $L$-$b/a$-controlled sample.
We adopt 3 bins for $b/a$ with edges at (0,0.5,0.7,0.8) and 4 bins
for \lrad with edges at ($10^{24}$,$10^{24.5}$,$10^{25}$,$10^{25.5}$,$10^{26}$) (W\,Hz$^{-1}$) to construct $3\times4$ cells in the $b/a$-\lrad space.
We construct an $L$-$b/a$-controlled sample by resampling 1200 AGNs from the smaller size sample while ensuring the source frequency of the control sample in each $b/a$-\lrad cell is the same as that of the larger size sample.
In this case, the $L$-$b/a$-controlled smaller size sample have similar \lrad-$b/a$ distribution as the larger size sample.
It is therefore expected that the LOS with respect to the minor axis of the galaxies are similar in the two samples, assuming that the galaxies have similar intrinsic shape.

The dPA distribution of the $L$-$b/a$-controlled sample is shown in the right panel of Fig. \ref{fig:size}.
The control sample still has a weaker minor-axis alignment trend than the larger size sample.
We repeat the sampling 100 times and the K-S test results always show $p_{\rm null}$ to be much smaller than 1\%, which suggests that the dPA distributions of the control sample and the larger size sample cannot be drawn from the same population.

In summary, we observed a stronger minor-axis alignment trend for AGNs with larger projected linear sizes.
This is not due to the dPA-\lrad relation.
We tried to use the $L$-$b/a$-controlled sample to take the projection effect into account to explain the size-dPA dependence, but the minor-axis alignment trend is still larger for AGNs with larger sizes.
We note that the $L$-$b/a$-controlled sample analysis is not enough to eliminate the projection effect. 
In Sect. \ref{app:proj}, we showed that a simple toy model assuming all jets have the same intrinsic length could produce a similar size-dPA dependence.
The projection effect in even the toy model cannot be assessed properly through $b/a$-controlled sample.
Therefore, we suggest that the size-dPA dependence observed in this work is, if not purely, largely caused by the projection effect.
To further assess the projection effect, it would be necessary to use a complete radio sample and have a deeper understanding on the radio emitting regions of radio AGNs, which is out of the scope of this work.  

       \begin{figure}
        \centering
        \includegraphics[width=\linewidth]{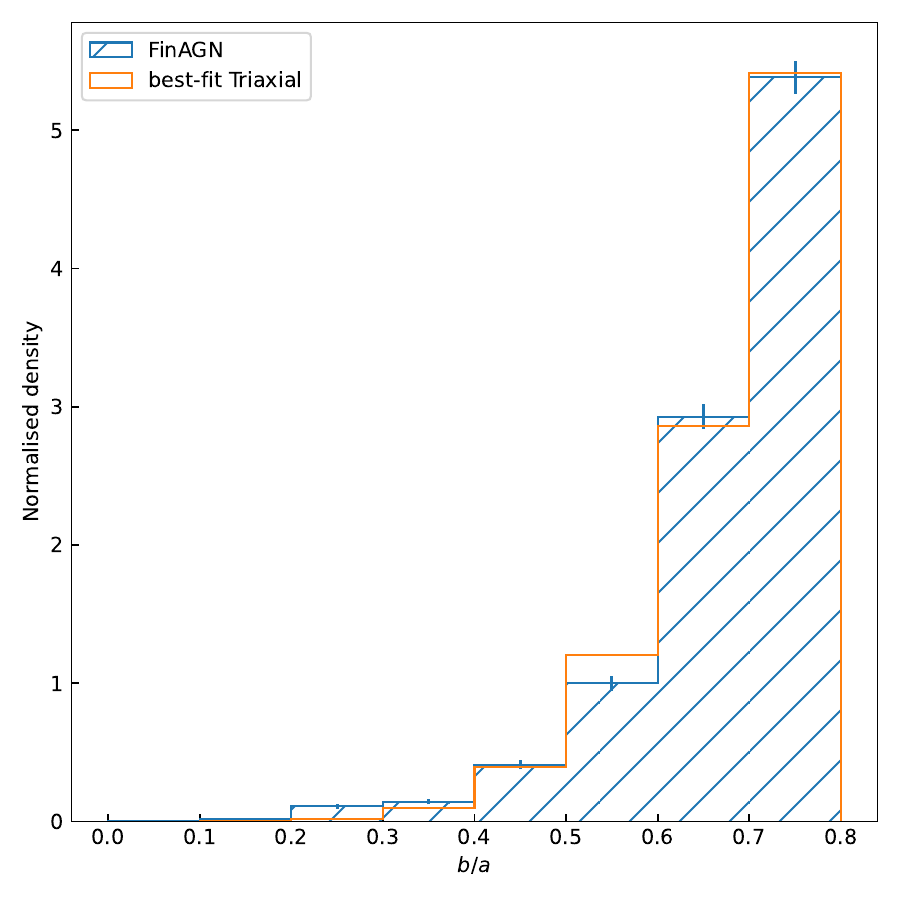}
        \caption{
        The normalised apparent axis ratio distributions for the FinAGN sample and the best-fit triaxial model.
        The blue hatch histogram denotes the FinAGN sample while the orange histogram denotes the triaxial model.}
        \label{fig:bafit}
    \end{figure}
\section{Apparent alignment to intrinsic alignment}\label{sec:intrinsic}
\label{atoi}

    The presented analyses are based on the projected alignment measured directly from radio and optical images. 
    In this section, we investigate the intrinsic AGN jet orientations based on some simple assumptions.
    
\subsection{Fitting axis ratio distribution}
    
    To produce a simulation of the projection effect, we need to model the real spatial orientations of radio AGN jets in relation to their host galaxies.
    The intrinsic shape of galaxies can be estimated based on the apparent optical axis ratio distribution \citep[e.g.][]{Sansom87,Ryden92,Fasano10,Chang13,Li18,Zhang22}.
    We follow their method to estimate the galaxy shape distributions statistically in a simple way. 
    We assume that the entire galaxy population can be divided into two different groups based on their shapes, i.e. oblate galaxies and triaxial galaxies \citep{Chang13}. 
    Oblate galaxies have two equal-sized axes and a shorter $z$-axis, corresponding to galaxies with a prominent disc component. 
    Triaxial galaxies have three axes with different sizes, corresponding to elliptical galaxies. 
    A sample with only triaxial galaxies or oblate galaxies results in different apparent axis ratio distributions. 
    The axis ratio distribution of a galaxy sample can be decomposed into an oblate and a triaxial component, thus providing information about the distribution of the intrinsic shape of the galaxies in the sample.
    The fraction of oblate galaxies and triaxial galaxies can then be used to reproduce the apparent dPA distributions. 
    
    To simulate the axis ratio of triaxial and oblate galaxies, we follow the recipes in \citet{Chang13} to simulate the oblate and triaxial populations, assuming the three axes of a galaxy to be $a\geq b\geq c$.
    The oblate galaxies are taken to have a mean intrinsic axis ratio $\epsilon=c/b=c/a=0.29$ and a standard deviation of $\sigma_\epsilon=0.07$. 
    The triaxial galaxies are described by two parameters, $T(=[1-\beta^2]/[1-\gamma^2])$ and ${E}(=1-\gamma)$ where ${\beta={b/a}}$ and ${\gamma={c/a}}$. The $T$ parameter follows a Gaussian distribution with a mean of ${T}=0.64$ and a standard deviation of ${\sigma_T=0.08}$. The $E$ parameter has a mean value of $E=0.41$ and a standard deviation of ${\sigma_E=0.19}$.
\begin{figure*}
    \centering
    \includegraphics[width=\textwidth]{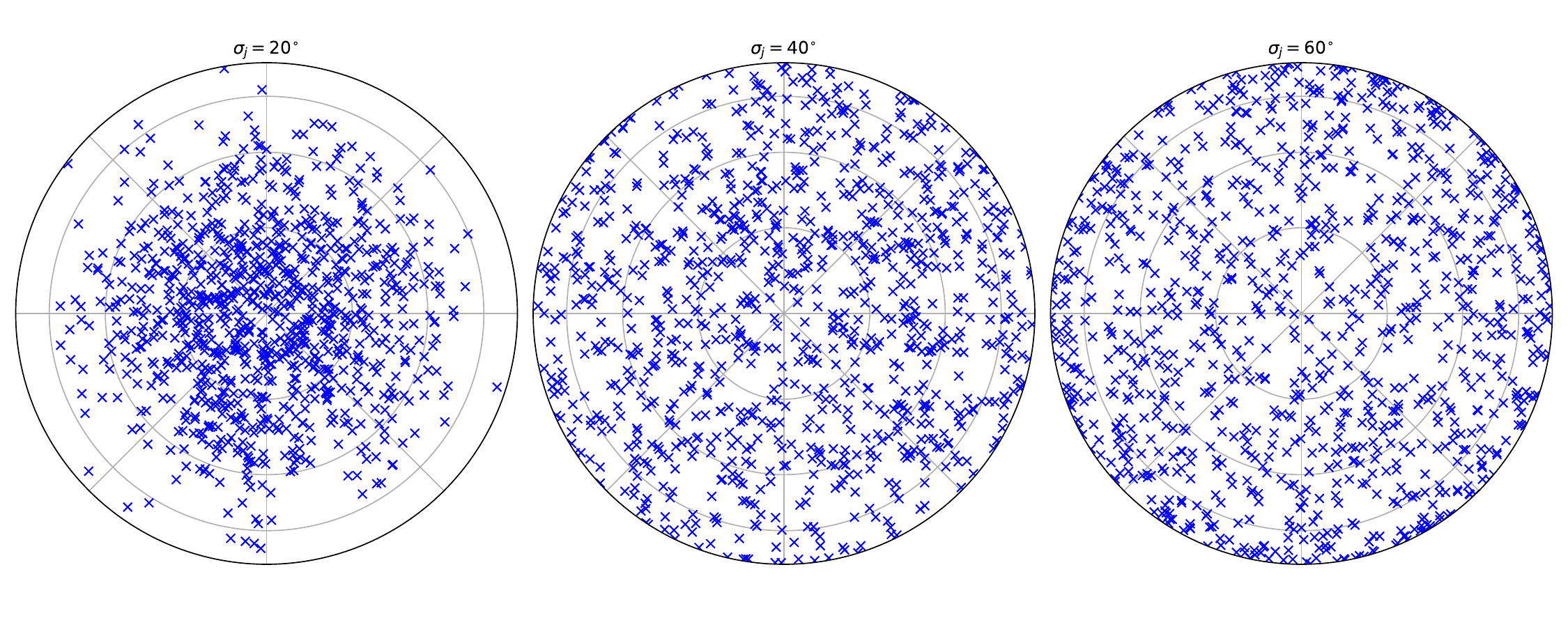}
    \caption{The distributions of the simulated jet pointings following Eq. \ref{eq:jetp} with different $\sigma_j$, 20\degree, 40\degree and 60\degree.
    The LOS in each plot is along the minor axis of the galaxy, i.e. the crosses at the center correspond to jet pointing exactly towards the minor axis, while the crosses at the edge correspond to jet pointing perpendicular to the minor axis.
    The $\sigma_j$ used in each panel is listed on top of the panel.
    Each plot contains 1\,000 different jet pointings.
    The inner big circles in each plots show angles of 20\degree, 40\degree and 60\degree with respect to the minor axis.
    }
    \label{fig:gausmodel}
\end{figure*}
    \begin{figure}
    \centering
    \includegraphics[width=\linewidth]{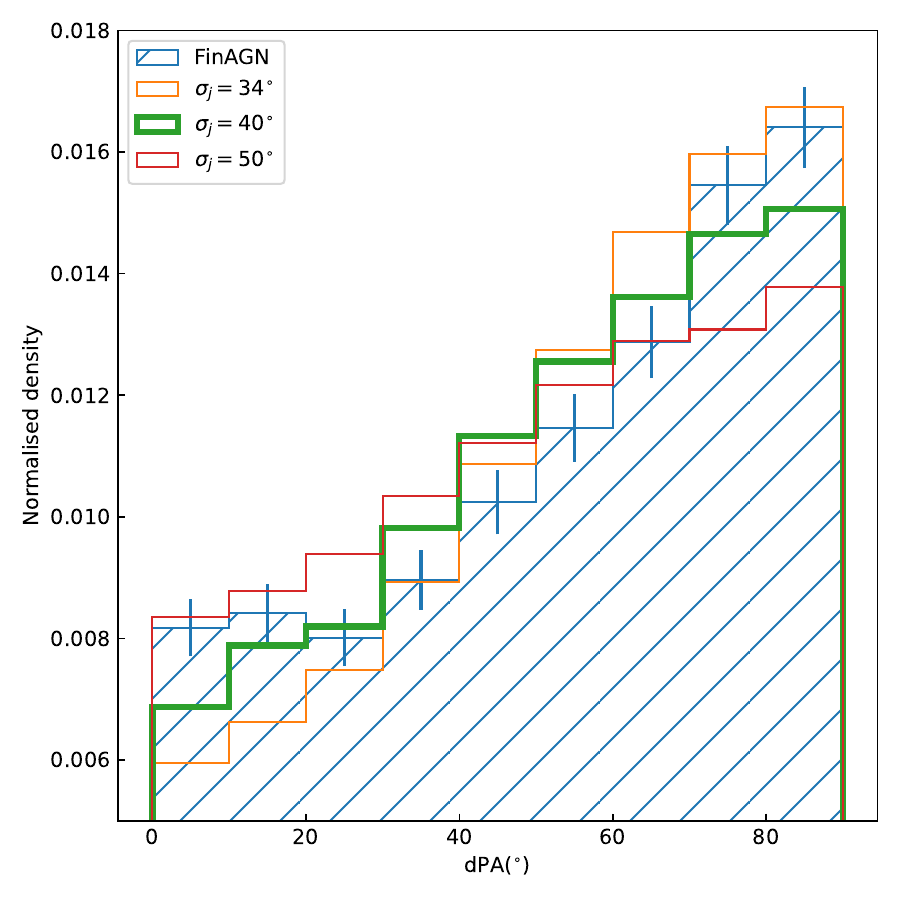}
    \caption{
    The simulated dPA distributions compared with the dPA distribution of the FinAGN sample.
    The hatched histogram with errorbars denotes the FinAGN sample.
    The $\sigma_j$ for three simulated distributions are listed in the figure.
    The K-S tests indicate a largest $p_{\rm null}$ corresponding to the smallest probability that the simulated and observed distributions are drawn from different populations when $\sigma_j=46$\degree, which is marked with thick lines in the figure.
    }
    \label{fig:5}       
    \end{figure}

    We define the galaxy's frame $(x, y, z)$ with the $x$-axis along the major axis of the galaxy and $z$-axis along the minor axis, and an observer's frame $(x', y', z')$ with $z'$-axis pointing from the centre of the galaxy to the observer and $x'$-axis in the $xy$ plane.
    For an observer at a line of sight (LOS) with polar coordinates $(\theta, \phi)$ in the galaxy's frame, \citet{Binney85} gave the equation for the projected ellipse as 
    \begin{equation}\label{eq:ellip}
        R^2 = Ax'^2+Bx'y'+Cy'^2
    \end{equation}
    where $R$ is a radius-related constant and 
    
    \begin{equation} \label{eq:A}
      {{A} \equiv \frac{{\rm cos}^2\theta}{\beta^2}({\rm sin}^2\phi+\frac{{\rm cos}^2\phi}{\gamma^2})+\frac{{\rm sin}^2\theta}{\gamma^2}} \,,
    \end{equation} 
    
    \begin{equation} \label{eq:B}
      {{B}  \equiv  {\rm cos}\theta {\rm sin} 2\phi (1-\frac{1}{\gamma^2})\frac{1}{\beta^2}} \,,
    \end{equation} 
    
    \begin{equation} \label{eq:C}
      {{C}  \equiv  (\frac{{\rm sin}^2\phi}{\gamma^2}+{\rm cos}^2\phi)\frac{1}{\beta^2}} \,.
    \end{equation}
    
    The projected apparent axis ratio $q$ of the galaxy is then,  
    
   \begin{equation}
      {1\geq{{q}(\theta, \phi, \beta, \gamma)=\sqrt{\frac{A+C-\sqrt{(A-C)^2+B^2}}{A+C+\sqrt{(A-C)^2+B^2}}}}} \,,
    \end{equation} 

    To model the axis ratio distribution, we created a sample of 10\,000 galaxies and simulated apparent axis ratio distributions for both the oblate and the triaxial populations, defining the fraction of oblate galaxies in a sample to be ${f_{\rm ob}}$. 
    We combined two axis ratio distributions with different $f_{\rm ob}$ and compared them with the observed axis ratio distributions. 
    Additionally, the fit was performed using only the observed/simulated apparent axis ratio between 0 and 0.8 to match the source selection processes in Sect. \ref{sec:opa}. 

    We found the best-fit oblate fraction for the FinAGN sample with the minimum ${\chi^2}$ statistics is 0, indicating a single triaxial component in the axis ratio distribution.
    This is in line with the implications in \citet{Barisic19} and \cite{Zheng20} that powerful radio AGNs are mostly triaxial galaxies.    

    As the fitting result indicated a pure triaxial population for the FinAGN sample, we varied the $T$, $E$, $\sigma_T$ and $\sigma_E$ with a step of 0.02 and generated triaxial models to fit the axis ratio distribution of the FinAGN sample.
    We found the best-fit $T$ and $E$ to be 0.84 and 0.2 respectively while the $\sigma_T$ and $\sigma_E$ were not changed.
    The modelled and observed axis ratio distributions are shown in Fig. \ref{fig:bafit}.

\begin{figure*}
    \centering
    \includegraphics[width=0.66\textwidth]{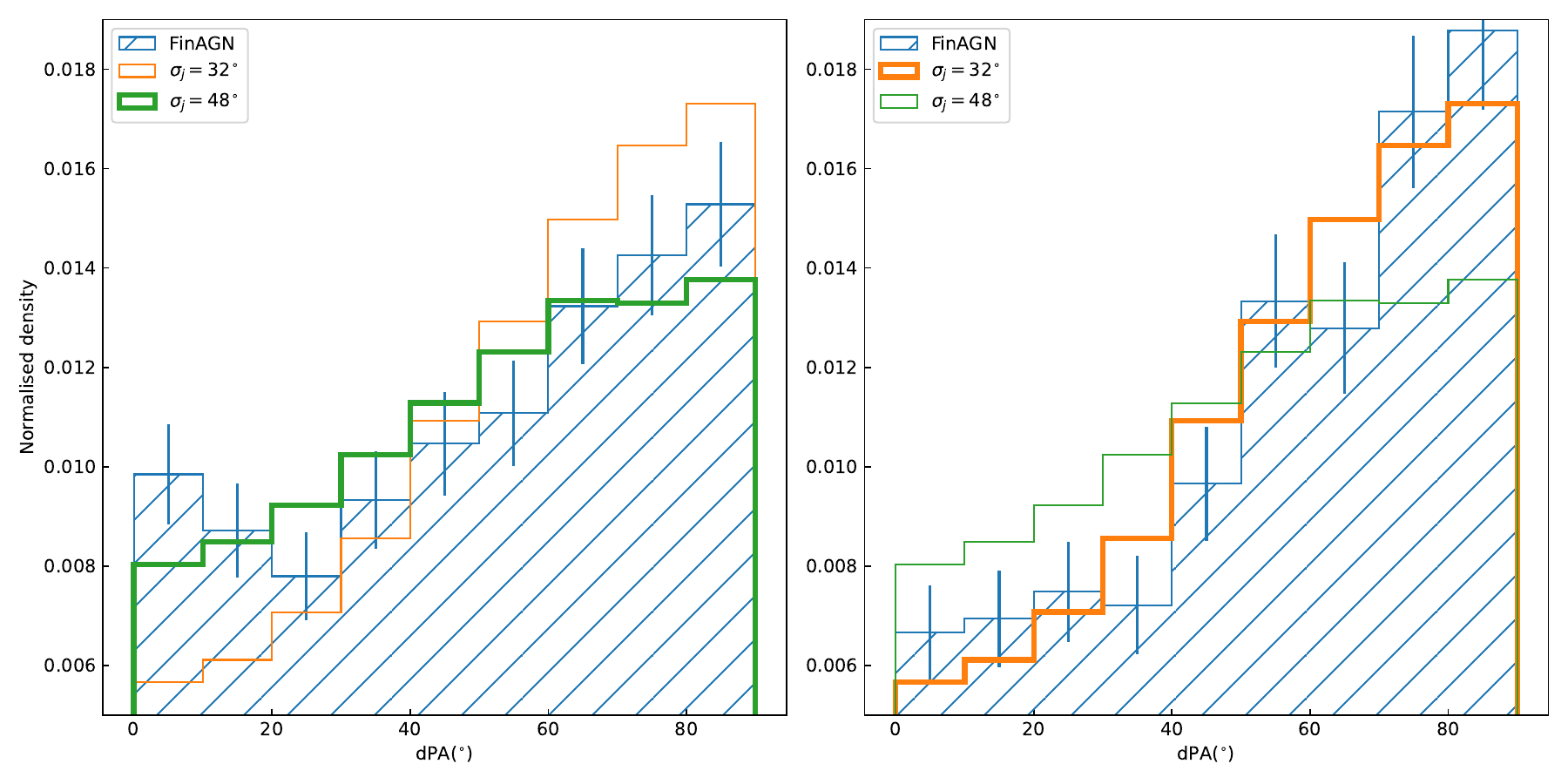}
    \caption{
    (From left to right) The simulated dPA distributions compared with the dPA distribution of the moreL\&massive and the lessL\&massive sample.
    The hatched histograms with errorbars denote the observed distributions in each panel.
    The $\sigma_j$ for the two simulated distributions are listed in the two panels.
    The simulated distribution corresponding to the largest $p_{\rm null}$ in each sample is highlighted in each panel.
    }
    \label{fig:simobs_bm} 
\end{figure*}
\subsection{Simulation of the dPA distribution}

    While measuring the intrinsic jet orientations in a large sample from imaging is impossible, we can study it statistically by reproducing the apparent alignment distributions with simple modelled intrinsic orientations.
    For given LOS angles $(\theta, \phi)$, we first calculated the PA of the apparent major axis of the projected galaxy, which has an intrinsic $T$, $E$ as the best-fit model in the previous section.
    The PA of the apparent major axis can be derived from Eq. \ref{eq:ellip} as described in \citet{Binney85}\footnote{We note that the PA of the major axis calculated in this way does not follow a uniform distribution because the real north-south direction for an observer on earth is not taken into account in the calculation. This corresponds to a rotation along the $z'$-axis of the observer's frame (a different $x'$-axis). However, such rotation would be applied to both the apparent major axis and the apparent jet and the final dPA would not change.}. 
    For a jet pointing from the galactic centre to direction $(\theta_j,\phi_j)$, where the polar coordinates are in the galaxy's frame, the apparent PA can be obtained by projection from the galaxy's frame to the $x'y'$ plane.
    In this way, we can obtain the apparent dPA for a jet observed from a given LOS.

    We assumed the intrinsic offset between the jet and the minor axis of the galaxy $\theta_j$ follows a Gaussian distribution with a standard deviation of $\sigma_j$ at a fixed $\phi_j$ defined within 0\degree to 90\degree.
    Then the overall probability density function of the $\theta_j$ is 
    \begin{equation}\label{eq:jetp}
        p_{\rm gaus}(\theta_j|\sigma_j)= N\,{\rm sin}(\theta_j)\frac{{\rm exp}({\frac{\theta_j^2}{2\sigma_j^2}})}{\sigma_j},\,\,\, (0\leq\theta_j\leq90^{\circ}).
    \end{equation}
    where $N$ is the normalisation factor,
    \[
    N = 1/\int_{0}^{90}{\rm sin}(\theta)\frac{{\rm exp}({\frac{\theta^2}{2\sigma_j^2}})}{\sigma_j} {\rm d}\theta 
    \]
The $\sigma_j$ can be used to estimate the level of the minor-axis alignment. 
As shown in Fig.\ref{fig:gausmodel}, a smaller $\sigma_j$ suggests more jets are close to the minor axis.

For each $\sigma_j$ from 30\degree to 60\degree with a step of 2\degree, we generated 100\,000 galaxies with $T$ and $E$ following a best-fit distribution, 100\,000 different jet angles following the distribution of Eq. \ref{eq:jetp}, and 100\,000 uniformly distributed LOS angles\footnote{$p(\theta)\propto{\rm sin}(\theta)$ and $p(\phi)=\rm Constant$}.
To match the data selection processes for the observed data, we excluded the simulated sources with an apparent axis ratio larger than 0.8.
Finally, we obtained the simulated dPA distributions for each $\sigma_j$.
    
\subsection{Comparing Model to Observation}
\begin{figure*}
    \centering
    \includegraphics[width=\textwidth]{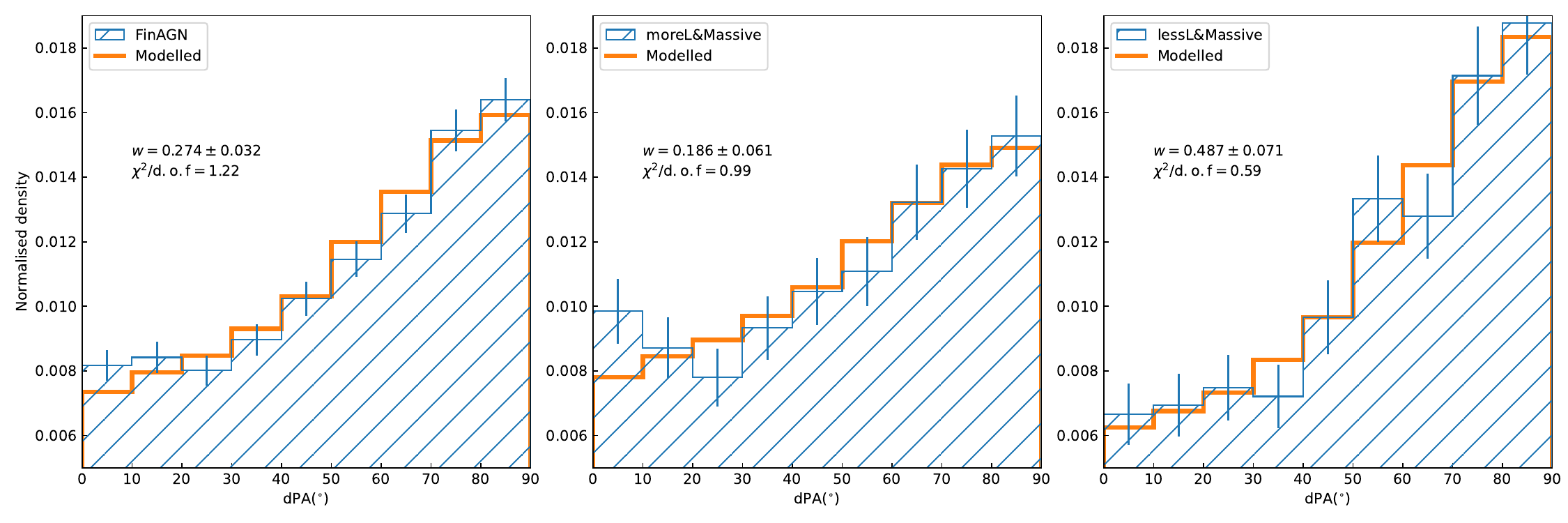}
    \caption{
     (From left to right) The two-component simulated dPA distributions compared with the dPA distribution of the FinAGN, the moreL\&massive and the lessL\&massive sample.
    The hatched histograms with errorbars denote the observed distributions in each panel.
    The best-fit simulated distributions are highlighted with thick lines in all panels.
    The best-fit parameter $w$ is the fraction of the component with $\sigma_j=20$\degree.
    }
    \label{fig:2gaus} 
\end{figure*}
\begin{figure*}
    \centering
    \includegraphics[width=0.66\textwidth]{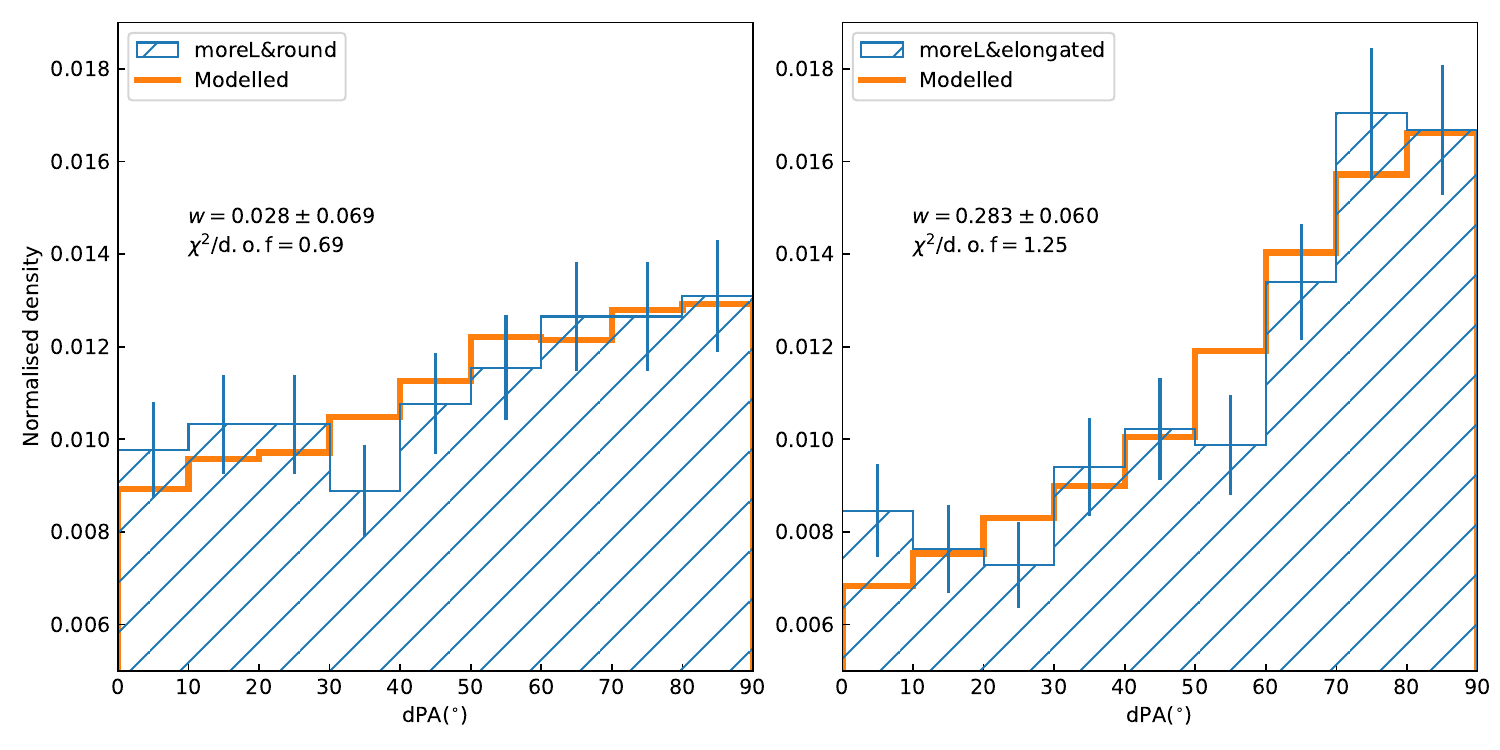}
    \caption{
    (From left to right) The two-component simulated dPA distributions compared with the dPA distribution of the moreL\&round and the moreL\&elongated sample defined in Sect. \ref{sec:Lmor}.
    The hatched histograms with errorbars denote the observed distributions in each panel.
    The best-fit simulated distributions are highlighted with thick lines in all panels.
    The best-fit parameter $w$ is the fraction of the component with $\sigma_j=20$\degree.
    }
    \label{fig:2gaus_round} 
\end{figure*}
To find a jet distribution model that matches the FinAGN sample, we calculated the K-S statistics using the dPAs of the FinAGN sample and the modelled dPAs with different $\sigma_j$.
The K-S results showed that the largest $p_{\rm null}$ is 5.4\%, corresponding to a $\sigma_j=40$\degree.
This means that at $\sigma_j=40$\degree, the fitting results gives the smallest probability that the simulated and observed distributions are drawn from different populations.
We show the dPA distribution with $\sigma_j=40$\degree along with the dPA distribution of the FinAGN sample in Fig. \ref{fig:5}.
To show the change of dPA distributions with $\sigma_j$, we also plotted two dPA distributions with $\sigma_j=34$\degree and 50\degree, which have $p_{\rm null}\ll 1\%$.
We also compared the simulated dPA distributions with the moreL\&massive and lessL\&massive samples in Sect. \ref{sec:LM}.
The K-S tests indicate a largest $p_{\rm null}=31.2\%$ at $\sigma_j=48$\degree for the moreL\&massive sample and a largest $p_{\rm null}=50.1\%$ at $\sigma_j=32$\degree for the lessL\&massive sample.
These results are shown in Fig. \ref{fig:simobs_bm}.
We note that the models with $\sigma_j=42-52$\degree all have a $p_{\rm null}>5\%$ for the moreL\&massive sample.
For the lessL\&massive sample, the models with $\sigma_j=30-36$\degree all have a $p_{\rm null}>5\%$.
This result suggests that radio jets with higher luminosity in general have a larger offset with respect to the minor axis of the host galaxies.

We note that although the K-S results suggest no significant difference between the dPA distributions of the observed samples and some of the simulated dPA distributions, they do not indicate that these models fit well.
As shown in Fig.\ref{fig:5} and Fig.\ref{fig:simobs_bm}, the simulated dPA distributions with the largest $p_{\rm null}$ can be outside the 1-$\sigma$ limit in the histograms.
We calculate the reduced $\chi^2$ for these dPA distributions with 20\degree bins, and none of the results for the FinAGN sample indicate a good fit ($\chi^2$/d.o.f close to 1, where d.o.f is the degree of freedom in the fitting).
This indicates that a single Gaussianised $\theta_j$ distribution is not adequate to describe the apparent dPA distributions.

To better mimic the intrinsic jet alignment, we use a two-component model to describe the $\theta_j$ distribution,
\begin{equation}
    p(\theta_j|w) = w p_{\rm gaus}(\theta_j|20^{\circ})+(1-w) p_{\rm gaus}(\theta_j|60^{\circ})
\end{equation}
The $\theta_j$ distributions for the two components $p_{\rm gaus}(\theta_j|20^{\circ})$ and $p_{\rm gaus}(\theta_j|60^{\circ})$ are shown in the first and the third panels of Fig. \ref{fig:gausmodel}.
The first component represents a population in which the jets are more likely to be aligned to the minor axis of the galaxy, while the second component represents a population with more randomly pointing jets.
Note that the choice of 20\degree and 60\degree is based on the need to have a minor-axis aligned component and a less aligned component: one component should have a larger than $\sigma_j$ than the best-fit $\sigma_j$ in the single component fitting ($\sim40$\degree) and the other component should have a smaller $\sigma_j$.
Other combinations of angles might also work as long as they can provide fits that are as good as the model in this work.
In this model, a larger $w$ means a stronger minor-axis alignment tendency.

We used the least-square method to find the best-fit weights $w$ of the first component for the FinAGN sample, the moreL\&massive and the lessL\&massive samples.
The best-fit results are shown in Fig. \ref{fig:2gaus} and Tab. \ref{tab:2gaus}.
We can see that the two-component models fit the observed dPA distributions well, with all reduced $\chi^2$ close to one and $0.16<P_{\rm null}<0.84$\footnote{The range of 0.16 to 0.84 corresponds to the 1-$\sigma$ confidence interval in the Gaussian distribution.}, which indicates a good fit and no over-fitting.

The simulation results show that the $w$ of the moreL\&massive and lessL\&massive samples are significantly different.
This confirms the conclusion based on the K-S test in Sect. \ref{sec:LM} and implies that the jet alignment depends on the radio luminosity in our work.
Radio AGNs with a lower luminosity have jets more likely to be aligned with the minor axis of the host galaxies.

We also selected the simulated sources with an apparent axis ratio of $b/a>0.7$ and $b/a\leq0.7$ as the simulated round and elongated sources.
They were used to fit the moreL\&round and the moreL\&elongated samples defined in Sect. \ref{sec:Lmor}.
The best-fit results are shown in Fig. \ref{fig:2gaus_round} and Tab. \ref{tab:2gaus}.
The two-component model fits the moreL\&elongated sample well.
The small best-fit $w$ for the moreL\&round sample implies that the jets in these radio AGNs are less likely to be aligned with the minor-axis of the host galaxies than those in the other samples.
The non-overlapping confidence intervals of the best-fit $w$ for the moreL\&round and moreL\&elongated samples suggests that the flattened dPA distribution in the moreL\&round sample cannot be explained by the projection effect.
Therefore, the jet alignment is also dependent on the shape of galaxies.

As the simulation results could depend on the intrinsic galaxy morphology model, i.e. $T$ and $E$, we repeated the simulations with $T$ and $E$ 0.1 larger or smaller than the best-fit $T$, $E$.
Although the projected axis ratio of these galaxy morphology models are significantly different from the observed distribution, the best fit $w$ do not change significantly and most of the resulting $w$ fall in the 1-$\sigma$ ranges of the results using the best-fit galaxy morphology model, but with worse $P_{\rm null}$.
Therefore, our results seem robust and not very sensitive to the galaxy morphology model.

It should be noted that the small peak for the moreL\&Massive sample below dPA=30\% is not reproduced in our simple fitting. 
The peak indicates a 10\% to 20\% excess compared with the model distribution at dPA $<10\%$. This might imply an extra component in the samples.

\begin{table}
    \centering
    \begin{tabular}[width=\linewidth]{c|ccc}
    Sample     & $w$ & $\chi^2/d.o.f$ & $P_{\rm null}$\\
    \hline
    FinAGN     & 0.274$\pm0.032$ & 1.22 & 0.29 \\ 
    moreL\&massive & 0.186$\pm0.061$ & 0.99 & 0.44 \\
    lessL\&massive & 0.487$\pm0.071$ & 0.59 & 0.77 \\
    moreL\&round & 0.028$\pm0.069$ & 0.69 & 0.68 \\
    moreL\&elongated & 0.283$\pm0.06$ & 1.25 & 0.27\\
    \end{tabular}
    \caption{The best-fit results of the two-component fitting for different radio AGN samples. 
    The goodness-of-fit can be implied from the null-probability $P_{\rm null}$, which quantifies the probability that the simulated distribution is similar to the observation.}
    \label{tab:2gaus}
\end{table}

\subsection{Projection effect in size-dPA relation}\label{app:proj}
Here we use a simple toy model for the projection effect that might lead to the observed size-dPA relation discussed in Sect. \ref{sec:psize}.
We assume that at these low frequencies the luminosity of the extended radio sources is not a function of orientation.
In this case, the projected linear size is only a function of the angle between the jet and the LOS. 

We set an intrinsic jet orientation distribution with $\sigma_j=40$\degree and simulated the apparent dPA 100\,000 times following the processes described in Sect. \ref{sec:intrinsic}.
This time we also calculated the projected linear size based on the jet angle with respect to the LOS of each simulated source.
The resulting simulated sample was divided into two equal-size subsamples based on the projected sizes.

\begin{figure}
    \centering
    \includegraphics[width=\linewidth]{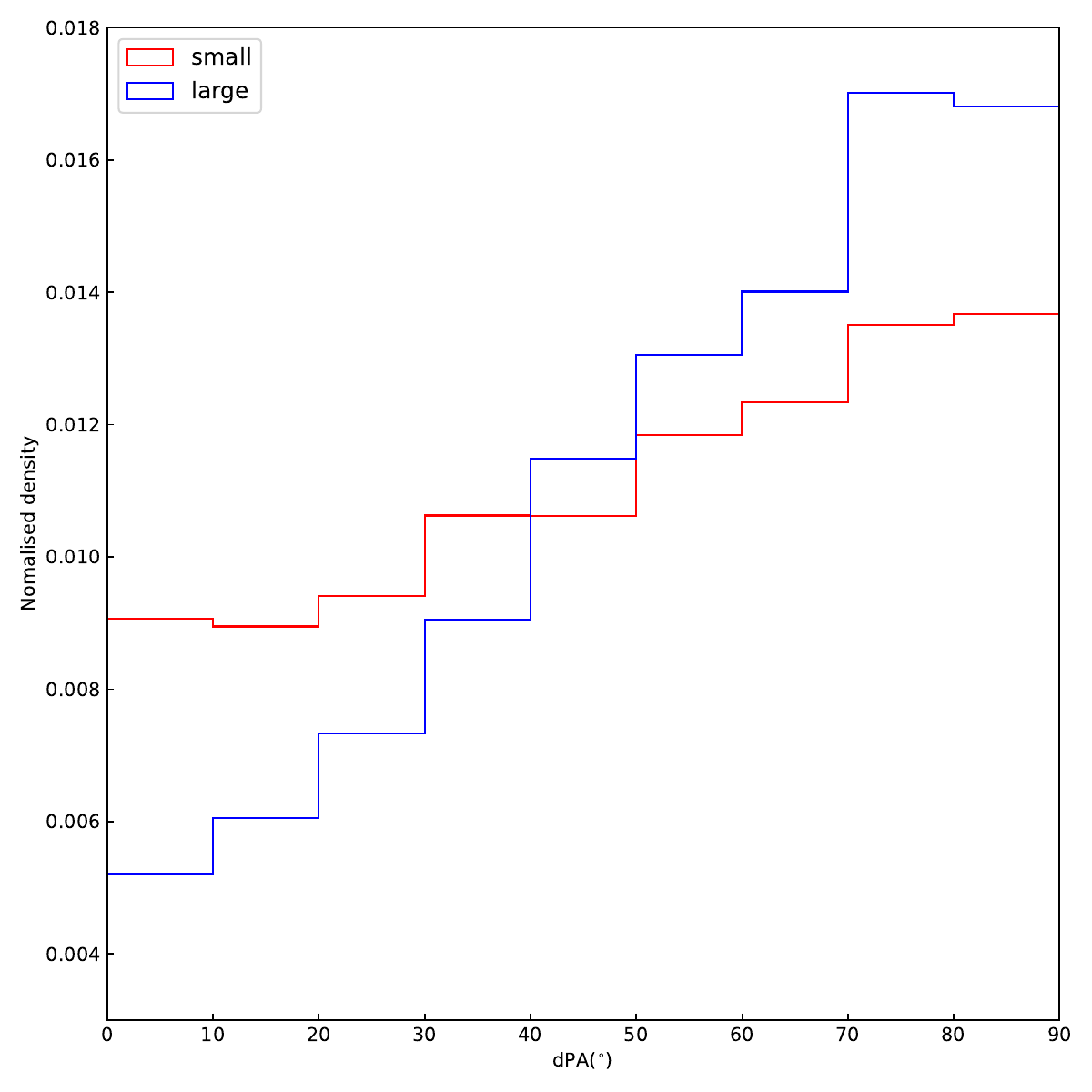}
    \caption{The simulated dPA distributions of two subsamples with different projected linear sizes in a toy model.
    The red histogram denotes the simulated subsample with smaller projected sizes.
    The blue histogram represents the $b/a$-controlled sample with larger projected sizes.
    }
    \label{fig:toy} 
\end{figure}

Similar to the process in Sect. \ref{sec:psize}, we resampled 10000 sources from the subsample with larger projected linear sizes to make a $b/a$-controlled sample.
This $b/a$-controlled sample has a $b/a$ distribution similar to that of the subsample with smaller sizes.
The dPA distributions of the subsample with smaller sizes and the $b/a$-controlled sample are shown in Fig. \ref{fig:toy}.

Apparently the $b/a$-controlled sample still shows a significantly stronger minor-axis aligned trend than the small-size subsample.
We stress that our toy model does not take into account many important factors in reality such as the resolution of surveys, the dependence (or bias) of size on luminosity, the distribution of the intrinsic size of radio sources.
However, such toy model still reproduced a similar size-dPA dependence as we observed. 

\section{Conclusions \& Discussion}
\label{dis}

    We have selected a radio AGN sample with reliable radio and optical position angle measurements, RPA and OPA, based on the LoTSS DR2 and FIRST data.
    We investigated the alignment result between radio-AGN jets relative to their host galaxies by comparing optical and radio images of radio galaxies, followed by modelling intrinsic jet alignment to apparent jet alignment. 
    Our conclusions are summarised as follows:
\begin{itemize}
    \item[i] In all the radio AGN samples, the dPA distributions indicate a minor-axis alignment, in contrast to the major-axis alignment distribution for the SF galaxies. 
    \item[ii] We divided the radio AGN sample into subsamples with different radio luminosity \lrad\, and stellar mass \mstar. Less luminous radio AGNs tend to have a stronger trend of minor-axis alignment. We found a possible up-turn or flattening trend at small dPAs ($\lesssim30$\degree) in all subsamples except the one with higher radio luminosity and lower stellar mass which is also the smallest subsample..
    \item[iii] We found a significant projected dPA difference between the optically round and more elongated hosts of radio AGNs. The more elongated radio AGNs tend to be more minor-axis aligned. This effect is not due to the fact that the OPA for rounder galaxies is more poorly defined.
    \item[iv] We found the radio AGNs with larger projected linear size tends to have a stronger minor-axis alignment trend than the AGNs with smaller size but similar \lrad\, and optical axis ratio. This is likely to be a result of projection effects.
    \item[v] We used a single-component and a two-component model to simulate the observed dPA distributions, where the components differ in the spread ($\sigma_j$) of intrinsic PA difference between the optical minor axis and the radio jet. The dPA distributions are best described by the two-component models. The fitting results indicate that the jet alignment is dependent on radio luminosity and the shape of galaxies, with the jets being more likely to be aligned with the minor-axis of the galaxy for lower radio luminosity and optically more elongated radio AGNs.
\end{itemize}

    The implication of the conclusions is that most of the SMBHs associated with these radio-AGN sources undergo a coherent accretion dominated process, as described in \citet{2011MNRAS.414.2148L}. Conversely, for SMBHs of moreL radio-AGN within massive host galaxies, their accretion processes can be related to the chaotic accretion or merger events \citep{2011MNRAS.414.2148L,2012MNRAS.425.1121H}. 
    Therefore their dPA distributions indicate that the radio-AGN jets are less aligned with the minor axis of the host galaxy.
    This scenario can also be implied from the less aligned dPA distribution for the round galaxies, which are more likely to have been through a major merger \citep{Conselice14,Somerville15,Cappellari16}.

    The minor-axis alignment between radio jets and their host galaxies in the radio AGN samples provides some important insights into the evolutionary history of galaxies. 
    As indicated by previous studies \citep{2017MNRAS.470.1559S,2018MNRAS.476.2801M,2019MNRAS.489.4016S,2020MNRAS.494.5713M,2021MNRAS.507.3985S}, in general, SMBH growth is dominated by a secular process resulting from gas inflow in both merger-free and merger-frequent galaxies \citep{1998ApJ...506L..97N}. 
    An important feature of the secular growth of SMBHs at centres of galaxies is the production of radio jets aligned with their host galaxies' angular momentum \citep{Beckmann22}. 
    Our result that the majority of radio jets are roughly aligned with their optical minor axis, therefore, reinforces this aforementioned evidence. 
    That is, the majority of galactic SMBHs grow via the coherent secular process between merger events, and associated SMBH spins result in radio-AGN jets mostly aligned toward their opitcal minor axis. 

    By contrast, for an extreme system, such as moreL radio-AGN hosted by massive galaxies, radio-AGN jets might not be aligned towards their optical minor axis. 
    The potential cause of the random orientations among these radio-AGN jets would be past merger events. 
    After a merger event, SMBH spins could be misaligned with respect to their host galaxies' angular momentum, and the spins might take time to realign under the Bardeen-Petterson effect \citep{1975ApJ...195L..65B}. 
    The timescale for this effect is thought to be much shorter compared to the secular process of SMBH growth \citep{1998ApJ...506L..97N}. 
    Hence, the misalignment is only observed in a small population of galaxies under extreme post-merger environments. 
    Another contributing factor for this is that our sample consists of mostly gas-poor early-type galaxies.
    For these galaxies after recent merger events, achieving a stable accretion inflow is thought to be unlikely, and chaotic accretion is a reasonable scenario, which prolongs post-merger misalignments in SMBH jets \citep{Bustamante19}.
    
    Our results provide essential observational evidence for the simulation results in recent studies \citep[e.g.][]{Beckmann22}, especially regarding the alignment between SMBH spins and their host galaxies' angular momentum. 
    Provided the evidence of SMBH growth dominated by coherent secular process in long-lasting epochs between merger events, additional evidence of the influence of radio-AGN outflows on their host galaxies might be crucial to further understand the overall co-evolution between galaxies and their central SMBHs.

\begin{acknowledgements}
We thank the anonymous referee for the very useful suggestions to improve this paper.
XCZ acknowledges support from the CSC (China Scholarship Council)-Leiden University joint scholarship program.
XCZ acknowledges support from the National SKA Program of China (Grant No. 2022SKA0120102).
LOFAR data products were provided by the LOFAR Surveys Key Science project (LSKSP; https://lofar-surveys.org/) and were derived from observations with the International LOFAR Telescope (ILT). LOFAR \citep{vHaarlem13} is the Low Frequency Array designed and constructed by ASTRON. It has observing, data processing, and data storage facilities in several countries, that are owned by various parties (each with their own funding sources), and that are collectively operated by the ILT foundation under a joint scientific policy. The efforts of the LSKSP have benefited from funding from the European Research Council, NOVA, NWO, CNRS-INSU, the SURF Co-operative, the UK Science and Technology Funding Council and the Jülich Supercomputing Centre. 

This research uses services or data provided by the Astro Data Lab at NSF's National Optical-Infrared Astronomy Research Laboratory. NOIRLab is operated by the Association of Universities for Research in Astronomy (AURA), Inc. under a cooperative agreement with the National Science Foundation.

The Legacy Surveys consist of three individual and complementary projects: the Dark Energy Camera Legacy Survey (DECaLS; NOAO Proposal ID \# 2014B-0404; PIs: David Schlegel and Arjun Dey), the Beijing-Arizona Sky Survey (BASS; NOAO Proposal ID \# 2015A-0801; PIs: Zhou Xu and Xiaohui Fan), and the Mayall z-band Legacy Survey (MzLS; NOAO Proposal ID \# 2016A-0453; PI: Arjun Dey). DECaLS, BASS and MzLS together include data obtained, respectively, at the Blanco telescope, Cerro Tololo Inter-American Observatory, NSF's National Optical Infrared Astronomy Research Laboratory (NOIRLab); the Bok telescope, Steward Observatory, University of Arizona; and the Mayall telescope, Kitt Peak National Observatory, NOIRLab. The Legacy Surveys project is honored to be permitted to conduct astronomical research on Iolkam Du’ag (Kitt Peak), a mountain with particular significance to the Tohono O’odham Nation.

The NSF’s NOIRLab is operated by the Association of Universities for Research in Astronomy (AURA) under a cooperative agreement with the National Science Foundation. Database access and other data services are provided by the Astro Data Lab.

BASS is a key project of the Telescope Access Program (TAP), which has been funded by the National Astronomical Observatories of China, the Chinese Academy of Sciences (the Strategic Priority Research Program "The Emergence of Cosmological Structures" Grant \# XDB09000000), and the Special Fund for Astronomy from the Ministry of Finance. The BASS is also supported by the External Cooperation Program of Chinese Academy of Sciences (Grant \# 114A11KYSB20160057), and Chinese National Natural Science Foundation (Grant \# 11433005).

The Legacy Surveys team makes use of data products from the Near-Earth Object Wide-field Infrared Survey Explorer (NEOWISE), which is a project of the Jet Propulsion Laboratory/California Institute of Technology. NEOWISE is funded by the National Aeronautics and Space Administration.

The Legacy Surveys imaging of the DESI footprint is supported by the Director, Office of Science, Office of High Energy Physics of the U.S. Department of Energy under Contract No. DE-AC02-05CH1123, by the National Energy Research Scientific Computing Center, a DOE Office of Science User Facility under the same contract; and by the U.S. National Science Foundation, Division of Astronomical Sciences under Contract No.AST-0950945 to NOAO.

This project used data obtained with the Dark Energy Camera (DECam), which was constructed by the Dark Energy Survey (DES) collaboration. Funding for the DES Projects has been provided by the U.S. Department of Energy, the U.S. National Science Foundation, the Ministry of Science and Education of Spain, the Science and Technology Facilities Council of the United Kingdom, the Higher Education Funding Council for England, the National Center for Supercomputing Applications at the University of Illinois at Urbana-Champaign, the Kavli Institute of Cosmological Physics at the University of Chicago, Center for Cosmology and Astro-Particle Physics at the Ohio State University, the Mitchell Institute for Fundamental Physics and Astronomy at Texas A\&M University, Financiadora de Estudos e Projetos, Fundação Carlos Chagas Filho de Amparo, Financiadora de Estudos e Projetos, Fundação Carlos Chagas Filho de Amparo à Pesquisa do Estado do Rio de Janeiro, Conselho Nacional de Desenvolvimento Científico e Tecnológico and the Ministério da Ciência, Tecnologia e Inovação, the Deutsche Forschungsgemeinschaft and the Collaborating Institutions in the Dark Energy Survey. The Collaborating Institutions are Argonne National Laboratory, the University of California at Santa Cruz, the University of Cambridge, Centro de Investigaciones Enérgeticas, Medioambientales y Tecnológicas–Madrid, the University of Chicago, University College London, the DES-Brazil Consortium, the University of Edinburgh, the Eidgenössische Technische Hochschule (ETH) Zürich, Fermi National Accelerator Laboratory, the University of Illinois at Urbana-Champaign, the Institut de Ciències de l'Espai (IEEC/CSIC), the Institut de Física d'Altes Energies, Lawrence Berkeley National Laboratory, the Ludwig-Maximilians Universität München and the associated Excellence Cluster Universe, the University of Michigan, the National Optical Astronomy Observatory, the University of Nottingham, the Ohio State University, the University of Pennsylvania, the University of Portsmouth, SLAC National Accelerator Laboratory, Stanford University, the University of Sussex, and Texas A\&M University.

We thank Emma Rigby for polishing the text.
\end{acknowledgements}

\bibliographystyle{aa} 
\bibliography{cite.bib} 

\begin{appendix}

\section{FinAGN information table}
We provide the information of the FinAGN sample as online materials for people interested in further investigation.
Tab. \ref{tab:finagn} shows the column descriptions of the online table.
More related information can be found by matching the source name and IDs with the LoTSS DR2 VAC, the Legacy Surveys and the SDSS database.

\begin{table*}[t!]
    \centering
    \begin{tabularx}{\linewidth}{l|X}
    \hline\hspace{2em}
        Columns &  Descriptions \\\hline
        Source\_Name & The unique ID for radio sources from the LoTSS DR2. Sources without a LoTSS DR2 ID are named based on the sky coordinates in the SDSS.\\
        ls\_id$^{\dagger}$ & The unique ID in the Legacy Surveys. \\
        plate\_sdss & SDSS plate ID if a SDSS counterpart is present, else (-1). \\
        mjd\_sdss & SDSS MJD ID. \\
        fiber\_sdss & SDSS Fiber ID. \\
        ra & The right ascension (J2000) of the optical counterpart. \\
        dec & The declination (J2000) of the optical counterpart. \\
        z & The best redshift measurement. \\
        z\_source & The source of the redshift. `SDSS' if taken from the SDSS; `DESI' if taken from the DESI spectroscopic survey; `HETDEX' if from the HETDEX; `Phot' if it is a photo-z from \citet{Duncan22}.\\
        radioPA & Position angle of the radio source, defined from north to east.\\
        optPA & Position angle of the optical source, defined from north to east.\\
        dPA & The radio-optical misalignment angle.\\
        logM & The log($M_{\star}/M_{\odot}$) of the galaxy. If the source is in the FIRSTAGN sample, the stellar mass is based on the MPA-JHU results, else estimated from infrared luminosity as in \citet{Wen13}. \\
        logL150 & The log(\lrad/$(\rm W\,Hz^{-1})$) estimated in this work. \\
        b/a & The opitical axis ratio. If the source is in the FIRSTAGN sample, it is `deVAB\_r' in the SDSS, else estimated from the Legacy Surveys. \\
        AngSize & The projected largest angular size of the radio source in units of arcsecond.\\
        Size & The projected linear size of the radio source in units of kpc.\\
        Cat\_source & The source catalogue of which the radio source is from. `LoTSS' if it is from the LoTSS DR2 VAC, otherwise `FIRST'. Sources identified as GRGs in the \citet{Oei23} are `GRG'. \\\hline
        
    \end{tabularx}
    \caption{The column descriptions for the online FinAGN information catalogue.\\
    $^{\dagger}$: The ls\_id is different from the `Legacy\_ID' in the LoTSS DR2 VAC \citep{Hardcastle23}, but the same as in the Legacy Surveys.}
    \label{tab:finagn}
\end{table*}
\end{appendix}

\end{document}